\documentclass[preprint,showpacs,preprintnumbers,amsmath,amssymb]{revtex4}

\usepackage{graphicx}
\usepackage{dcolumn}
\usepackage{bm}
\usepackage{theorem}
\newcommand{\mib}[1]{\mbox{\boldmath $#1$}}

\def\N{{\bf N}}
\def\R{{\bf R}}
\def\W{{\bf W}}
\def\C{\bf C}

\def\0{\mib{0}}

\def\B{{\bf B}}

\def\r{{\bf r}}
\def\x{{\bf x}}
\def\y{{\bf y}}

\def\n{{\bf n}}
\def\m{{\bf m}}

\def\vlambda{\mib{\lambda}}

\def\P{{\bf P}}

\begin{document}

\preprint{}

\title{Maximum distributions of
bridges of noncolliding Brownian paths}

\author{Naoki Kobayashi}
\email{knaoki@phys.chuo-u.ac.jp}
\author{Minami Izumi}%
\email{izumi@phys.chuo-u.ac.jp}
\author{Makoto Katori}
\email{katori@phys.chuo-u.ac.jp}
\affiliation{%
Department of Physics,
Faculty of Science and Engineering,
Chuo University, 
Kasuga, Bunkyo-ku, Tokyo 112-8551, Japan 
}%

\date{27 August 2008}

\begin{abstract}
The one-dimensional Brownian motion
starting from the origin at time $t=0$,
conditioned to return to the origin at time $t=1$
and to stay positive during time interval
$0 < t < 1$, is called the Bessel bridge
with duration 1.
We consider the $N$-particle system
of such Bessel bridges conditioned never to collide
with each other in $0 < t < 1$,
which is the continuum limit
of the vicious walk model in watermelon
configuration with a wall.
Distributions of maximum-values of paths attained in the
time interval $t \in (0,1)$ are studied
to characterize the statistics of
random patterns of the repulsive paths
on the spatio-temporal plane.
For the outermost path, the distribution function
of maximum value is exactly 
determined for general $N$.
We show that the present $N$-path system of noncolliding
Bessel bridges is realized as the positive-eigenvalue
process of the $2N \times 2N$ matrix-valued
Brownian bridge in the symmetry class C.
Using this fact computer simulations are performed 
and numerical results on the $N$-dependence
of the maximum-value distributions
of the inner paths are reported.
The present work demonstrates that
the extreme-value problems of noncolliding paths
are related with
the random matrix theory, representation theory
of symmetry, and the number theory.

\end{abstract}

\pacs{05.40.-a,02.50.-r}

\maketitle

\section{Introduction}

Random walk (RW) models are important in physics,
chemistry, and computer sciences.
They can be used effectively, when we explain
basic concepts of statistical physics \cite{Rei65},
stochastic processes in physics and chemistry \cite{vKam92},
and stochastic algorithms \cite{MR95}.
In particular, RWs have been used to provide
simple and useful models to discuss 
various physical phenomena in far-from-equilibrium,
such as interface dynamics \cite{BS95,KSOYHM05,YSKOM07},
polymer networks \cite{dGen79,dGen68,EG95}, 
wetting and melting transitions \cite{Fis84}, and so on.
If we take the proper spatio-temporal continuum limit,
called the {\it diffusion scaling limit}, of the RW models,
the Brownian motion (BM) models are obtained.
By virtue of the limit procedure,
the BM models are enriched with mathematics.
The following example of a conditional BM 
will demonstrate this statement \cite{Yor95, BPY01, KIK08}.

\begin{figure}
\includegraphics[width=0.6\linewidth]{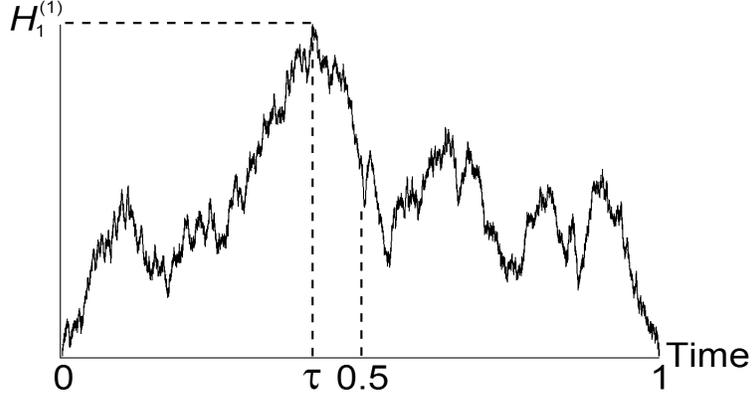}
\caption{A sample of path of the 3-dimensional Bessel bridge
with duration 1. 
In this example the time $\tau$, at which the height
of the bridge attains its maximum, is less than 1/2.}
\label{fig:Fig1}
\end{figure}

Consider the one-dimensional standard BM, $B(t), t \geq 0$,
where $\langle B(t) \rangle=0$ and
$\langle B(s) B(t) \rangle = \min\{s,t\}$.
The BM conditioned to stay positive $B(t) > 0, 0 < t < \infty$,
is called the {\it 3-dimensional Bessel process},
abbreviated as BES(3),
since it is equivalent in distribution with the radial part of
the 3-dimensional BM and its transition
probability density is given by a special case
of the modified Bessel function
(see Sec.II.A below and
3.3 C in \cite{KS91}, VI.3 in \cite{RY98}, 
IV.34 in \cite{BS02}).
When we consider the case that it starts from 
the origin at time $t=0$ and returns to the 
origin at time $t=1$, this conditional BES(3) is called
the {\it 3-dimensional Bessel bridge with duration 1}
starting from 0, since as illustrated by Fig.\ref{fig:Fig1},
the path drawn on the (1+1)-dimensional spatio-temporal
plane seems to be a bridge over the time axis.
(Note that by the scaling property between space and time
of BM, no generality is lost by setting the time duration
be 1.)
Let $r(t), 0 \leq t \leq 1$, denote the
3-dimensional Bessel bridge.
By symmetry, we can expect that the height
of the bridge attains its 
maximum value with the highest probability at time $t=1/2$.
The probability density 
for $r(t) \in dx$ at time $t=1/2$
is readily calculated as
\begin{equation}
p_{\rm BESb}(1/2, x)
= 8 \sqrt{\frac{2}{\pi}}
x^2 e^{-2 x^2}, \quad
0 \leq x < \infty
\label{eqn:pBb1/2}
\end{equation}
(see below Eq.(\ref{eqn:probab})),
which gives the moments
$\langle (r(1/2))^s \rangle=2^{-s}(s+1)!!$
if $s$ is even and
$\langle (r(1/2))^s \rangle
=\{2/\sqrt{\pi}\} 2^{-s/2} ((s+1)/2)!$
if $s$ is odd.
The shape of the present bridge is, however,
randomly deformed, and as we can see in Fig.\ref{fig:Fig1}
the time $0 < \tau < 1$,
at which $r(t)$ attains its maximum, 
will fluctuate around the mean time 
$\langle \tau \rangle =1/2$. 
We define
$$
H^{(1)}_1= \max_{0 < t < 1} r(t)
=r(\tau).
$$
We can show that the probability density
for $H^{(1)}_1 \in d h$ is given as
\begin{equation}
p_{H^{(1)}_1}(h)=8 \sum_{n=1}^{\infty}
e^{-2 n^2 h^2} 
(4 n^4 h^3 - 3 n^2 h), 
\quad 0 \leq h < \infty,
\label{eqn:pH1}
\end{equation}
which gives the moments
$$
\langle (H^{(1)}_1)^s \rangle
= \frac{s(s-1)}{2^{s/2}}
\Gamma(s/2) \zeta(s),
\quad s=0,1,2, \dots
$$
with the gamma function 
$\Gamma(z)=\int_{0}^{\infty} e^{-u} u^{z-1} du$.
Here $\zeta(s)$ is Riemann's zeta function
$$
\zeta(s)=\sum_{n=1}^{\infty} 
\frac{1}{n^s}, 
$$
which is an important special function
in number theory \cite{BPY01,Yor97}.
In the present paper, 
as multivariate generalization of the Bessel bridge \cite{KT07b},
we study the $N$-path systems
of the 3-dimensional Bessel bridges with duration 1, 
conditioned never to collide with each other in $0 < t < 1$;
$\r^{(N)}(t)=(r^{(N)}_1(t), r^{(N)}_2(t), \dots, 
r^{(N)}_N(t)),
0 \leq t \leq 1$, with the conditions
$\r^{(N)}(0)=\r^{(N)}(1)=\0$ and
$0 < r^{(N)}_1(t) < r^{(N)}_2(t) < \cdots < r^{(N)}_N(t),
0 < t < 1$.

The systems of RWs with nonintersecting condition were
introduced by M. E. Fisher as mathematical models
to describe the wetting and the melting transitions
and named {\it vicious walk models} \cite{Fis84}.
They have been used not only to discuss 
dynamics of domain walls and melting of commensurate
surfaces \cite{HF84}, but also to study
polymer networks \cite{dGen68,EG95},
the related enumerative combinatorial
problems \cite{AME91,GOV98,KGV00}, 
and nonequilibrium critical phenomena \cite{CK03}.
In general the noncolliding diffusion particle systems
are obtained as the diffusion scaling limits
of the vicious RW models \cite{KT02,KT03,Gill03}.
In particular, the version of the vicious RW model,
whose continuum limit is the {\it noncolliding Bessel bridges},
$\r^{(N)}(t), 0 \leq t \leq 1$, is called
the {\it N-watermelons with a wall}
\cite{EG95,GOV98,KGV00,Gill03,Kra06}.

In the discrete mathematics the one-dimensional RWs
conditioned to visit only nonnegative sites
${\N}_0=\{0,1,2, \dots \}$ is called the Dyck paths
and the asymptotics of the average height of the Dyck
paths in the long-step limit was studied by
de Bruijn, Knuth and Rice \cite{dBKR72}.
Recently Fulmek generalized this classical result
by evaluating the asymptotics of the average height
of the 2-watermelons with a wall \cite{Ful07}.
In this calculation, he showed the fact that the 
number-theoretical special functions, such as
Jacobi's theta function and the double Dirichlet series
are useful to describe the asymptotics.
Motivated by this important observation,
the present authors \cite{KIK08} studied
the $N=2$ case of the noncolliding Bessel
bridges, $\r^{(2)}(t)=(r^{(2)}_1(t), r^{(2)}_2(t))$,
which is the continuum limit of the 2-watermelons with a wall,
and clarified that this phenomenon is indeed
the generalization of the relationship
between the maximum-value distribution of the
3-dimensional Bessel bridge and Riemann's zeta function
mentioned above.

We will report in this paper both of the exact
results and the numerical results on the maximum-value
distributions of $N$ paths in the noncolliding Bessel bridges.
Main exact result is the following determinantal
expression for the maximum-value distribution of the 
outermost path
({\it i.e.} the height of the continuum limit of watermelons),
\begin{equation}
H^{(N)}_N = \max_{0 < t < 1} r^{(N)}_N(t)
\label{eqn:HN1}
\end{equation}
for the $N$-path system
$\r^{(N)}(t)=(r^{(N)}_1(t), r^{(N)}_2(t), \dots, 
r^{(N)}_N(t))$;
\begin{equation}
\P(H^{(N)}_N < h)
= \frac{(-1)^N}{2^{N^2} \prod_{j=1}^{N} (2j-1)!}
\det_{1 \leq j, k \leq N}
\left[ \sum_{n=-\infty}^{\infty} H_{2(j+k-1)}
(\sqrt{2} n h) e^{-2 n^2 h^2} \right],
\label{eqn:Main1}
\end{equation}
where $H_n(x)$ denotes the $n$-th Hermite polynomial defined by
\begin{equation}
H_n(x)=n ! \sum_{k=0}^{[n/2]} 
\frac{(-1)^k (2x)^{n-2k}}{k ! (n-2k)!}
\label{eqn:Hermite1}
\end{equation}
with $[a]$ = the largest integer that is not
greater than $a$.
(We can see that, since $H_2(x)=4x^2-2$,
Eq.(\ref{eqn:Main1}) with $N=1$ and the
relation $p_{H^{(1)}_1}(h)=d \P(H^{(1)}_1 < h)/dh$
give the result (\ref{eqn:pH1}).)
The long-step asymptotics of all moments
of the height of the $N$-watermelons with a wall have been 
fully studied for arbitrary $N \geq 1$ by
Feierl \cite{Fei07,Fei08a}.
We will show that our result (\ref{eqn:Main1})
for the distribution functions of the continuous model
is consistent with
the results by Feierl for the moments of his
discrete model.
Quite recently Schehr {\it et al.} \cite{SMCRF08} 
showed their study
on the same problem and others by the
path-integral technique, 
a different method both from ours and Feierl's.
They also reported an expression
for the maximum-value distribution
(Eq.(5) in their paper \cite{SMCRF08}), which is different from
(\ref{eqn:Main1}).
We will show that the equivalence of these two
expressions is guaranteed
by the functional equation satisfied by
Jacobi's theta function $\vartheta_3(x,y)$.

\begin{figure}
\includegraphics[width=0.6\linewidth]{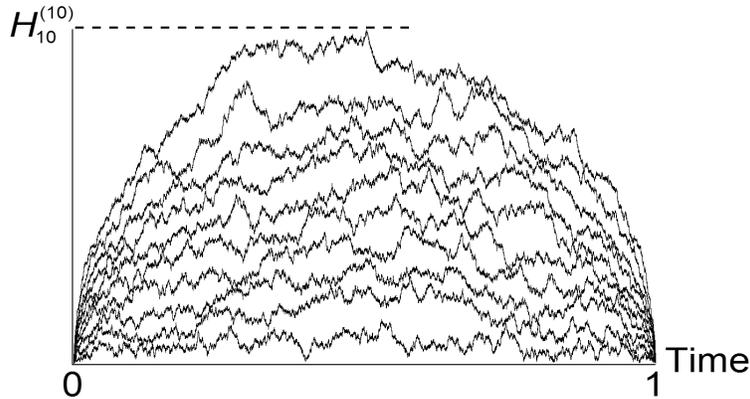}
\caption{A sample of paths of $N = 10$ noncolliding Bessel bridges 
with duration 1, all starting from 0 and
returning to 0, 
realized by the eigenvalue process.}
\label{fig:Fig2}
\end{figure}

Dyson introduced a matrix-valued BM,
$M(t)=(M_{jk}(t)), t \geq 0$, in the space
of Hermitian matrices.
If the matrix size is $N$, the $N$ diagonal
elements, $M_{jj}(t), 1 \leq j \leq N$,
the $N(N-1)/2$ real parts of the upper-triangle 
elements, ${\rm Re} (M_{jk}(t)), 1 \leq j < k \leq N$, and
the $N(N-1)/2$ imaginary parts of them, 
${\rm Im} (M_{jk}(t)), 1 \leq j < k \leq N$, 
are given by independent one-dimensional
standard BMs, the total number of which is $N^2$.
Dyson showed that the $N$ eigenvalues of $M(t)$
behaves as an interacting diffusion particle system
on the real axis $\R$, in which the long-ranged repulsive
forces work between any pair of particles
with the strength proportional to the inverse
of the distance of two particles \cite{Dys62}.
This eigenvalue process is called Dyson's BM model
and it has been proved to be equivalent in distribution
with the system of $N$-BMs conditioned
never to collide with each other \cite{Bia95,Gra99}.
The correspondence between eigenvalue processes
of matrix-valued diffusion processes and
noncolliding particle systems has been studied
\cite{Gra99,Bru89,Bru91,KO01,KT04}.
In the present paper we will use the fact that
the $N$-noncolliding Bessel bridges
can be realized as the positive-eigenvalue
process of the $2N \times 2N$ 
{\it matrix-valued Brownian bridge},
whose distribution at each time $0 < t < 1$
is related with the random matrix theory
\cite{Meh04} with a special symmetry called
the class C in \cite{AZ96,AZ97}.
Figure \ref{fig:Fig2} shows a sample of paths of the
$N=10$ noncolliding Bessel bridges realized
by this eigenvalue process.
One can imagine that it is very hard to simulate
such paths all starting from the origin and returning to
the origin with noncolliding condition by
direct computer simulation.
The present paper will demonstrate that
the relationship between the random matrix ensembles
and the noncolliding particle systems \cite{KT04}
provides a practical method to study such conditional 
processes effectively by computer simulations.
We will report the numerical evaluation
of the maximum-value distributions of not only
the outermost path $r^{(N)}_N(t)$,
but also inner paths
$r^{(N)}_k(t), k=1,2, \dots, N-1$.

Both from the view-points of
statistical physics and 
of random matrix theory \cite{Meh04,KT07b},
the study of $N \to \infty$ limit is interesting and
important for noncolliding paths
\cite{KT07a,TW07,Kuij08}.
For the average value of the maximum values of the
outermost path, we can read the
following behavior from
the numerical work by Bonichon and Mosbah 
for the watermelons with a wall \cite{BM03},
\begin{equation}
\langle H^{(N)}_N \rangle \simeq
\sqrt{1.67 N -0.06},
\quad N \to \infty.
\label{eqn:BM1}
\end{equation}
Recently, Schehr {\it et al.} \cite{SMCRF08} gave an argument
that the numerical data of Bonichon and Mosbah
(\ref{eqn:BM1}) shows only a pre-asymptotic
behavior in large $N$
and the true asymptotics should be
\begin{equation}
\langle H^{(N)}_N \rangle \simeq \sqrt{2N},
\quad N \to \infty.
\label{eqn:Schehr1}
\end{equation}
In the present paper, we report the numerical 
study of the $N$-dependence
of the maximum-value distributions systematically
not only for the outermost path but also for
all inner paths and discuss the $N \to \infty$
asymptotics based on our numerical data.

The paper is organized as follows.
In Sec.II.A, after giving brief explanations of
the 3-dimensional Bessel bridge
and the Karlin-McGregor formula \cite{KM59},
we define the $N$-noncolliding Bessel
bridges by giving the transition probability densities.
A matrix representation of the symmetry class C
is shown in Sec.II.B and the matrix-valued
Brownian bridge in the symmetry class C is introduced.
The equivalence in distribution between its 
positive-eigenvalue process and the noncolliding Bessel
bridges is then stated.
The problems studied in this paper is announced
in Sec.II.C.
In Sec.III.A the exact expression of the distribution
function of the maximum value for the outermost path,
which is the same as Eq.(5) in \cite{SMCRF08},
is derived by our method (Lemma 1).
This exact expression is then transformed into two kinds
of determinantal expressions (Proposition 2 and Theorem 4)
in Sec.III.B,
one of which is Eq.(7) given above.
The key lemma in the transformation is a set of
equalities between infinite series involving
the Hermite polynomials (Lemma 3)
derived from the functional equation
of Jacobi's theta function $\vartheta_3(x,y)$.
The numerical study is reported in Sec.IV.
Concluding remarks are given in Sec.V.
Appendices are used to give proofs of some formulas.

\section{Models and Problems}
\subsection{Transition probability density of 
noncolliding Bessel bridges}

Let $B(t), t \geq 0$, be the one-dimensional standard BM
starting from the origin; $B(0)=0$.
For any series of times $t_0 \equiv 0 < t_1 < t_2 < \cdots < t_M,
M=1,2, \dots$, the probability that the BM stays in 
interval $[a_m, b_m]$ at each time $t_m, m=1,2, \dots, M$,
is given by
\begin{eqnarray}
&& \P \Big(B(t_m) \in [a_m, b_m], m=1,2, \dots, M \Big)
\nonumber\\
&=& \int_{a_1}^{b_1} dx_1 \int_{a_2}^{b_2} d x_2
\cdots \int_{a_M}^{b_M} dx_M
\prod_{m=0}^{M-1} p(t_{m+1}-t_m, x_{m+1}|x_m),
\nonumber
\end{eqnarray}
where the transition probability density
$p(t, y|x)$ from the position $x$ to $y$ during time 
period $t$ is given by the probability 
density of the normal distribution
with mean 0 and variance $t$,
$$
p(t, y|x)=\frac{1}{\sqrt{2 \pi t}}
\exp \left\{ - \frac{(y-x)^2}{2t} \right\},
\quad t>0, x, y \in \R.
$$
We consider the situation that an absorbing wall is
set at the origin and BM is absorbed if it arrives at the wall.
By the reflection principle of BM \cite{KS91},
the transition probability density of such an absorbing BM
is given by
\begin{eqnarray}
p_{\rm abs}(t, y|x) &=& p(t, y|x)-p(t,-y|x)
\nonumber\\
&=& \frac{1}{\sqrt{2 \pi t}}
\Big\{ e^{-(y-x)^2/(2t)} - e^{-(y+x)^2/(2t)} \Big\}
\label{eqn:pabs}
\end{eqnarray}
for $x, y > 0, t > 0$.
The survival probability of the absorbing BM starting from
$x>0$ for the time period $T$ is then given by
$$
{\cal N}(T,x)=\int_{0}^{\infty} p_{\rm abs}(T, y|x) dy,
$$
whose asymptotics in $x/\sqrt{T} \to 0$ is easily
evaluated as
$$
{\cal N}(T,x) \simeq \sqrt{\frac{2}{\pi}}
\frac{x}{\sqrt{T}} \quad \mbox{in} \quad
\frac{x}{\sqrt{T}} \to 0.
$$
The transition probability density of the BM
under the condition that it stays forever 
in the positive region $\R_{+} = \{x \in \R: x > 0\}$
is then given by
\begin{eqnarray}
p_{\rm BES(3)}(t, y|x)
&\equiv& \lim_{T \to \infty}
\frac{{\cal N}(T-t, y) p_{\rm abs}(t,y|x)}
{{\cal N}(T, x)} \nonumber\\
&=& \frac{y}{x} p_{\rm abs}(t, y|x)
\label{eqn:pBES3}
\end{eqnarray}
for $x>0, y \geq 0, 0 < t < \infty$.
The diffusion process whose transition probability density
is given by (\ref{eqn:pBES3}) is called the
3-dimensional Bessel process,
abbreviated as BES(3), by the following reasons.
Consider the $d$-dimensional BM,
$\B^{(d)}(t)=(B_1(t), B_2(t), \dots, B_d(t))$,
whose coordinates are given by independent one-dimensional
standard BMs $\{B_j(t)\}_{j=1}^{d}$.
The distance from the origin of the $d$-dimensional BM
\begin{eqnarray}
R^{(d)}(t) &=& |\B^{(d)}(t)|
\nonumber\\
&=& \sqrt{ (B_1(t))^2+(B_2(t))^2+ \cdots
+ (B_d(t))^2}
\nonumber
\end{eqnarray}
can be regarded as a diffusion process in
$\R_+ \cup \{0\}$ and its transition 
probability density is given by
\begin{equation}
p_{{\rm BES}(d)}(t, y|x)=
\frac{y^{\nu+1}}{x^{\nu}} \frac{1}{t}
e^{-(x^2+y^2)/(2t)} I_{\nu} \left(\frac{xy}{t} \right)
\label{eqn:pBESd}
\end{equation}
for $x > 0, y \geq 0, t > 0$ with
$$
\nu=\frac{d-2}{2},
$$
where
$I_{\nu}(z) \equiv \sum_{n=0}^{\infty}(z/2)^{2n+\nu}/
\{\Gamma(n+1) \Gamma(\nu+n+1)\}$ is the 
modified Bessel function with the gamma function
$\Gamma(z)=\int_{0}^{\infty} e^{-u} u^{z-1} du$.
The process $R^{(d)}(t)$ is called the $d$-dimensional
Bessel process, BES($d$), \cite{KS91,RY98,BS02}. 
The transition probability density (\ref{eqn:pBES3})
of the BM conditioned to stay positive
is equal to (\ref{eqn:pBESd}) with
$d=3$, {\it i.e.}, $\nu=1/2$, since
$I_{1/2}(z)=(e^{z}-e^{-z})/\sqrt{2 \pi z}$.

Consider the space of all configurations
of $N$ particles in $\R_+$ with a fixed order
of positions,
$$
\W_N^{\rm C}=
\{\x = (x_1, x_2, \dots, x_N) \in \R_{+}^{N}:
x_1 < x_2 < \cdots < x_N \},
$$
which is called the Weyl chamber of type C$_{N}$
in the representation theory \cite{FH91}.
For $\x=(x_1, x_2, \dots, x_N), 
\y=(y_1, y_2, \dots, y_N) \in \W_N^{\rm C}, t > 0$,
consider the following determinant
$$
\det_{1 \leq j, k \leq N}
\Big[ p_{\rm BES(3)}(t, y_j|x_k) \Big]
= \prod_{j=1}^{N} \frac{y_j}{x_j}
\det_{1 \leq j, k \leq N}
\Big[ p_{\rm abs}(t, y_j|x_k) \Big],
$$
where the equality is given by the relation (\ref{eqn:pBES3})
and multilinearity of determinant.
By the theory of Karlin and McGregor \cite{KM59}
(see also \cite{KT05} with \cite{Lin73} and \cite{GV85}),
the probability that $N$ particle system of 
BES(3)'s starting from the 
configuration $\x \in \W_N^{\rm C}$ can keep the
order of $N$ particle positions by avoiding
any collision of particles 
for time period $T > 0$ is given by
$$
{\cal N}_N^{\rm C}(T, \x)
= \int_{0}^{\infty} dy_1 \cdots \int_{0}^{\infty} dy_N
\det_{1 \leq j, k \leq N}
\Big[ p_{\rm BES(3)} (T, y_j|x_k) \Big].
$$
By the Markov property of diffusion processes,
if we assume that the configuration at time $t=1$ is
fixed to be $\y \in \W_N^{\rm C}$,
for $0 < t_1 < t_2 < 1$,
the transition probability density from
$\x^{(1)} \in \W_N^{\rm C}$ at time $t_1$ to
$\x^{(2)} \in \W_N^{\rm C}$ at time $t_2$ is given as
\begin{eqnarray}
&& p_{\y}^{(N)}(t_1, \x^{(1)}; t_2, \x^{(2)})
\nonumber\\
&=& \frac{\displaystyle{\det_{1 \leq j, k \leq N}
\Big[p_{\rm BES(3)}(1-t_2, y_j|x^{(2)}_k) \Big]
\det_{1 \leq j, k \leq N}
\Big[p_{\rm BES(3)}(t_2-t_1, x^{(2)}_j|x^{(1)}_k) \Big]}}
{\displaystyle{\det_{1 \leq j, k \leq N}
\Big[p_{\rm BES(3)}(1-t_1, y_j|x^{(1)}_k) \Big]}}
\nonumber\\
&=& \frac{\displaystyle{\det_{1 \leq j, k \leq N}
\Big[p_{\rm abs}(1-t_2, y_j|x^{(2)}_k) \Big]
\det_{1 \leq j, k \leq N}
\Big[p_{\rm abs}(t_2-t_1, x^{(2)}_j|x^{(1)}_k) \Big]}}
{\displaystyle{\det_{1 \leq j, k \leq N}
\Big[p_{\rm abs}(1-t_1, y_j|x^{(1)}_k) \Big]}}.
\label{eqn:pNy}
\end{eqnarray}
Let $|\x|^2=\sum_{j=1}^{N} x_j^2$ and define
\begin{equation}
h_N^{\rm C}(\x)= \prod_{1 \leq j < k \leq N}
(x_k^2-x_j^2) \prod_{\ell=1}^{N} x_{\ell}
\label{eqn:hNC}
\end{equation}
for $\x=(x_1, x_2, \dots, x_N) \in \W_N^{\rm C}$.
Since we have known the asymptotics
$$
\det_{1 \leq j, k \leq N}
\Big[ p_{\rm abs}(t, y_j|x_k) \Big]
\simeq \frac{t^{-N(2N+1)/2} e^{-|\x|^2/(2t)} }
{2^{N(2N-1)/2} 
\prod_{j=1}^{N} \{ \Gamma(j) \Gamma(j+1/2) \} }
h_N^{\rm C}(\x) h_N^{\rm C}(\y)
$$
in $|\y| \to 0$
(the $\nu=1/2$ case of Eq.(33) in \cite{KT04}),
(\ref{eqn:pNy}) gives the following,
\begin{eqnarray}
&& p^{(N)}(t_1, \x^{(1)}; t_2, \x^{(2)})
\equiv \lim_{|\y| \to 0}
p^{(N)}_{\y}(t_1, \x^{(1)}; t_2, \x^{(2)})
\nonumber\\
&& \quad = \left( \frac{1-t_2}{1-t_1} \right)^{-N(2N+1)/2}
\frac{h_N^{\rm C}(\x^{(2)})}{h_N^{\rm C}(\x^{(1)})}
\nonumber\\
&& \qquad \times 
\det_{1 \leq j, k \leq N} \Big[
p_{\rm abs}(t_2-t_1, x^{(2)}_j|x^{(1)}_k) \Big]
\exp \left\{-\frac{|\x^{(2)}|^2}{2(1-t_2)}
+ \frac{|\x^{(1)}|^2}{2(1-t_1)} \right\}
\label{eqn:pN1}
\end{eqnarray}
for $\x^{(1)}, \x^{(2)} \in \W_N^{\rm C}$
and $0 < t_1 < t_2 < 1$, and
\begin{eqnarray}
&& p^{(N)}(0, \0; t, \x)
\equiv \lim_{|\x^{(1)}| \to 0}
p^{(N)}(0, \x^{(1)}; t, \x)
\nonumber\\
&& \quad = \frac{ \{t(1-t)\}^{-N(2N+1)/2} }
{(\pi/2)^{N/2} \prod_{j=1}^{N}(2j-1)!}
\Big\{ h_N^{\rm C}(\x) \Big\}^2
\exp \left\{- \frac{|\x|^2}{2t(1-t)} \right\}
\label{eqn:pN2}
\end{eqnarray}
for $\x \in \W_N^{\rm C}$ and $0 < t < 1$.
The $N$-particle system of 
noncolliding 3-dimensional Bessel bridges with duration 1
all starting from the origin is defined as the diffusion 
process such that its transition probability density
is given by (\ref{eqn:pN1}) and (\ref{eqn:pN2}),
and denoted by $\r^{(N)}(t)=(r^{(N)}_1(t), 
r^{(N)}_2(t), \dots, r^{(N)}_N(t)), 0 \leq t \leq 1$.
That is, for any sequence of times
$t_0 \equiv 0 < t_1 < t_2 < \cdots < t_M < 1,
M=1,2, \dots$, and
for any sequence of regions $\Delta_m \in \W_N^{\rm C},
m=1,2, \dots, M$,
\begin{eqnarray}
&& \P \Big( \r^{(N)}(t_m) \in \Delta_m,
m=1,2, \dots, M \Big)
\nonumber\\
&=& \int_{\Delta_1} d \x^{(1)}
\cdots \int_{\Delta_M} d \x^{(M)}
\prod_{m=0}^{M-1}
p^{(N)}(t_m, \x^{(m)}; t_{m+1}, \x^{(m+1)}).
\label{eqn:probab}
\end{eqnarray}
Note that no generality is lost by setting
the time duration be 1,
by the scaling property between space and time
of the present $N$-particle system
inherited from BM via BES(3).
From now on we call $\r^{(N)}(t), 0 \leq t \leq 1$,
simply the {\it $N$-noncolliding Bessel bridges}
for short.
Remark that, if we set $N=1$ and $t=1/2$
in (\ref{eqn:pN2}),
$h_1^{\rm C}(x)=x$ and Eq.(\ref{eqn:pBb1/2}) 
is obtained.

\subsection{Matrix-valued Brownian bridge in symmetry class C}

For $N \geq 1$, let $I_N$ be the $N \times N$ unit matrix
and define the $2N \times 2N$ matrix
$$
J=\left( \begin{array}{cc}
0_N & I_N \cr
-I_N & 0_N
\end{array} \right),
$$
where $0_N$ denotes the $N \times N$ zero-matrix.
Let ${\cal H}(N)$ and ${\cal S}(N; \C)$ be the collections
of all $N \times N$ Hermitian matrices and
all $N \times N$ complex symmetric matrices,
respectively.
Then consider the following set of $2N \times 2N$
Hermitian matrices,
\begin{equation}
{\cal H}^{\rm C}(2N)
=\left\{ C=\left( 
\begin{array}{cc}
H & S \cr S^{\dagger} & -{^{t}H}
\end{array} 
\right) ;
H \in {\cal H}(N), S \in {\cal S}(N; \C) 
\right\},
\label{eqn:HC}
\end{equation}
where ${^{t}H}$ denotes the transpose of $H$
and $S^{\dagger} \equiv {^{t}\overline{S}}$ denotes
the Hermitian conjugate of $S$.
We can see that any element $C \in {\cal H}^{\rm C}(2N)$ 
satisfies the relation
\begin{equation}
{^{t}C} J + J C = 0,
\label{eqn:symLie}
\end{equation}
which means that $C \in {\cal H}^{\rm C}(2N)
\subset {\cal H}(2N)$ satisfies
the symplectic Lie algebra; symbolically
written as
${\cal H}^{\rm C}(2N)={\rm sp}(2N, {\C}) \cap {\cal H}(2N)$
\cite{FH91}.
Due to the additional symmetry (\ref{eqn:symLie}),
the $2N$ real eigenvalues of $C \in {\cal H}^{\rm C}(2N)$
are given in the form
$\vlambda=(\lambda_1, \lambda_2, \dots, \lambda_N,
-\lambda_1, - \lambda_2, \dots, - \lambda_N)$,
where $\lambda_j \geq 0, 1 \leq j \leq N$.
Altland and Zirnbauer studied ${\cal H}^{\rm C}(2N)$
as the set of the Hamiltonians in the 
Bogoliubov-de Gennes formalism for the BCS 
mean-field theory of superconductivity,
with regarding the pairing of positive and
negative eigenvalues $(\lambda_j, - \lambda_j),
1 \leq j \leq N$, as the particle-hole symmetry
in the Bogoliubov-de Gennes theory \cite{AZ96,AZ97}.
They called ${\cal H}^{\rm C}(2N)$ 
(a representation of) the symmetry class C, since 
${\rm sp}(2N, \C)$ is denoted by ${\rm C}_N$ in 
Cartan's notation \cite{Hel78}.

The {\it Brownian bridge with duration 1} starting from
the origin
is defined as the one-dimensional standard BM
starting from 0 conditioned to return to 0 at time $t=1$
and denoted by $b(t), 0 \leq t \leq 1$.
The transition probability density of $b(t)$ is
given by
\begin{equation}
p_{\rm Bb}(s,x; t, y)
= \frac{p(1-t, 0|y) p(t-s, y|x)}{p(1-s, 0|x)}
\label{eqn:pBb}
\end{equation}
for $0 \leq s < t \leq 1, x, y \in \R$.
Let $b^{\rho}_{jk}(t), 0 \leq \rho \leq 2,
1 \leq j \leq k \leq N$, and
$\widetilde{b}^{0}_{jk}(t), 1 \leq j < k \leq N$, 
be independent Brownian bridges
with duration 1 starting from the origin.
Put
$$
s^{\rho}_{jk}(t)= \left\{
\begin{array}{cc}
b^{\rho}_{jk}(t)/\sqrt{2} & \mbox{if $j<k$} \cr
b^{\rho}_{jj}(t) & \mbox{if $j=k$} \cr
b^{\rho}_{kj}(t)/\sqrt{2} & \mbox{if $j > k$}
\end{array} \right.
$$
for $0 \leq \rho \leq 2$, and
$$
a^{0}_{jk}(t)= \left\{
\begin{array}{cc}
\widetilde{b}^{0}_{jk}(t)/\sqrt{2} & \mbox{if $j<k$} \cr
0 & \mbox{if $j=k$} \cr
-\widetilde{b}^{0}_{kj}(t)/\sqrt{2} & \mbox{if $j > k$,}
\end{array} \right.
$$
and consider the $N \times N$ matrices
$S^{\rho}(t)=\Big(S^{\rho}_{jk}(t) \Big)_{1 \leq j, k \leq N},
0 \leq \rho \leq 2$, and
$A^{0}(t)=\Big( a^{0}_{jk}(t) \Big)_{1 \leq j, k \leq N}$.
Then the $2N \times 2N$ matrix-valued BM is defined by
\begin{equation}
C^{(N)}(t)= \left( \begin{array}{cc}
S^{0}(t)+ i A^{0}(t) & S^{1}(t)+ i S^{2}(t) \cr
S^{1}(t)-i S^{2}(t) & -(S^{0}(t)-i A^{0}(t))
\end{array} \right),
\quad 0 \leq t \leq 1.
\label{eqn:Ct1}
\end{equation}
In order to define $C^{(N)}(t)$, we have used
$N(N+1)/2 \times 3 + N(N-1)/2=N(2N+1)$
independent Brownian bridges.
By definition (\ref{eqn:Ct1}),
$C^{(N)}(t) \in {\cal H}^{\rm C}(2N), 0 \leq t \leq 1$,
and $C^{(N)}(0)=C^{(N)}(1)=0_{2N}$.
That is, $C^{(N)}(t)$ can be regarded
as a Brownian bridge in the $N(2N+1)$-dimensional
space ${\cal H}^{\rm C}(2N)$.

\begin{figure}
\includegraphics[width=0.6\linewidth]{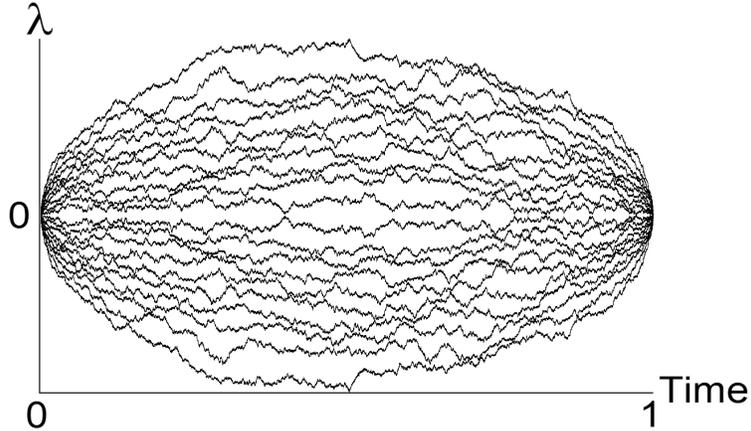}
\caption{A sample of paths of the eigenvalue process
$\vlambda^{(N)}(t), 0 \leq t \leq 1$, of the
matrix-valued Brownian bridge $C^{(N)}(t), 0 \leq t, \leq 1$,
in the symmetry class C for $N=10$.}
\label{fig:Fig3}
\end{figure}

At each time $0 < t < 1$, $C^{(N)}(t)$ can be diagonalized by 
a unitary-symplectic matrix and we can obtain
the eigenvalue process
$\vlambda^{(N)}(t)=(\lambda^{(N)}_1(t),
\dots, \lambda^{(N)}_N(t), -\lambda^{(N)}_1(t), 
\dots, -\lambda^{(N)}_N(t))$
with $0 \leq \lambda^{(N)}_1(t) \leq \cdots
\leq \lambda^{(N)}_N(t)$.
Using the generalized Bru's theorem given in \cite{KT04},
we can determine the transition probability density
for the positive part of eigenvalue process
$\vlambda^{(N)}_{+}(t)=(\lambda^{(N)}_1(t), \lambda^{(N)}_2(t), 
\dots, \lambda^{(N)}_N(t))$.
The result is exactly the same as Eqs.(\ref{eqn:pN1}) and
(\ref{eqn:pN2}).
It implies that $\vlambda_+^{(N)}(t) \in \W_N^{\rm C},
0 < t < 1$, with probability one, 
and the present positive-eigenvalue process
$\vlambda_{+}^{(N)}(t)$ is equivalent 
in distribution with the noncolliding Bessel
bridges $\r^{(N)}(t)$.
Figure \ref{fig:Fig3} shows a sample of paths of the eigenvalue
process $\vlambda^{(N)}(t), 0 \leq t \leq 1$,
for $N=10$ generated by computer
(see Sec.IV for detail).
There we can see ten positive paths
$\vlambda_{+}^{(N)}(t)$ and their
mirror images with respect to the line $\lambda=0$.
The sample of paths of the noncolliding Bessel bridges,
$\r^{(N)}(t), 0 \leq t \leq 1$,
shown in Fig.\ref{fig:Fig2} for $N=10$ is just obtained
by the upper half of this figure.

\subsection{Problems}
For the $N$-path system of noncolliding Bessel bridges,
$\r(t), 0 \leq t \leq 1$, we study the maximum
values for each path attained in the time interval
$t \in (0,1)$,
\begin{equation}
H^{(N)}_k \equiv \max_{0 < t < 1}
r^{(N)}_{k}(t), \quad 1 \leq k \leq N.
\label{eqn:HN}
\end{equation}
The problem considered in the present paper
is to clarify the statistical property
of the random variables $H^{(N)}_k, 
1 \leq k \leq N$.
We will report the exact expressions
of the distribution function 
for the outermost path $H^{(N)}_N$
for general $N \geq 1$
and the numerical results for inner paths.
The dependence of $N$ is studied
and the asymptotics in $N \to \infty$ will be
discussed.

\section{Exact Results for the Outermost Paths}
\subsection{Distribution function of $H^{(N)}_N$}

In this section we derive an exact expression
for the distribution function of 
the maximum value of the outermost path,
$\P(H^{(N)}_N < h)$.
In order to that first we consider the absorbing BM
in an interval $(0, h)$ for $h > 0$, in which
two absorbing walls are put at the origin and at the
position $x=h$.
The transition probability density
$p^{h}_{\rm abs}(t, y|x)$ for $t > 0, 
0 \leq x, y \leq h$ is the solution of the
diffusion equation
$\partial u(t,y)/\partial t
=(1/2) \partial^2 u(t,y)/\partial y^2$
with the initial condition
$
\lim_{t \to 0} u(t, y)=\delta(y-x)
$
and with the Dirichlet boundary conditions
$
u(t, 0)=u(t, h)=0, 
t \geq 0.
$
By the method of separation of variables
and the Fourier analysis, the unique solution 
is determined as
(see, for example, \cite{KGV03}),
\begin{equation}
p_{\rm abs}^{h}(t, y|x) = 
\frac{2}{h}\sum_{n = 1}^{\infty}
\exp
\left(-\frac{n^2 \pi^2}{2 h^2}t \right)
\sin \left(\frac{n\pi}{h}y \right)
\sin \left(\frac{n\pi}{h}x \right)
\label{eqn:phabs}
\end{equation}
for $t > 0, 0 \leq x, y \leq h$.
Note that it is the different expression of
the function
\begin{eqnarray}
p_{\rm abs}^{h}(t, y|x)
&=& \sum_{n=-\infty}^{\infty} \Big\{
p(t, y|x+2hn)- p(t, y|-x+2hn) \Big\}
\nonumber\\
&=& \frac{1}{\sqrt{2 \pi t}}
\sum_{n=-\infty}^{\infty} \left[
\exp\left\{ -\frac{1}{2t}
(y-(x+2hn))^2 \right\}
- \exp\left\{ -\frac{1}{2t}
(y-(-x+2hn))^2 \right\}
\right],
\nonumber
\label{eqn:p2}
\end{eqnarray}
which was used in our previous paper \cite{KIK08}.

Consider the following two determinantal functions,
\begin{equation}
q^{(N)}(t, \y|\x)
= \det_{1 \leq j, k \leq N}
\Big[ p_{\rm abs}(t, y_j|x_k) \Big],
\quad t > 0, \x, \y \in \W_N^{\rm C}
\label{eqn:qN}
\end{equation}
and
\begin{equation}
q^{(N)}_{h}(t, \y|\x)
= \det_{1 \leq j, k \leq N}
\Big[ p^{h}_{\rm abs}(t, y_j|x_k) \Big],
\quad t > 0, \x, \y \in \W_N^{h},
\label{eqn:qNh}
\end{equation}
where $\W_{N}^{h}=\{\x \in (0, h)^{N} :
x_1 < x_2 < \cdots < x_N \}$.
By the theory of Karlin and McGregor \cite{KM59},
$q^{(N)}(t, \y|\x)$ (resp. $q^{(N)}_h(t, \y|\x)$)
is the probability for the $N$-dimensional
absorbing BM starting from $\x \in \W_N^{\rm C}$
(resp. $\x \in \W_N^{h}$) at time $t=0$
to survive during time period $t$ by
avoiding any hitting with the absorbing boundaries
of $\W_N^{\rm C}$ (resp. $\W_N^{h}$),
and to arrive at $\y \in \W_N^{\rm C}$
(resp. $\y \in \W_N^{h}$) at time $t$.
Note that if we think that the $N$-dimensional
vector $\x=(x_1, x_2, \dots, x_N)$ represents
the positions of $N$ particles on $\R_+$,
a hitting with the boundary of $\W_N^{\rm C}$
means a hitting of the innermost particle $x_1$
with the origin or any collision between neighboring
particles $x_j=x_{j+1}, 1 \leq j \leq N-1$.
Similarly a hitting with the boundary of $\W_N^{h}$
means $x_1=0$, or any collision of particles,
or a hitting of the outer most particle
$x_N$ with the wall at $x=h$.

Consider the $N$-particle system of noncolliding 
BES(3),
$\widetilde{\R}^{(N)}(t)=(\widetilde{R}^{(N)}_1(t),
\cdots,
\widetilde{R}^{(N)}_N(t))$
starting from the configuration $\x \in \W_N^{h}$ at time $t=0$;
$\widetilde{\R}^{(N)}(0)=\x$, and arriving at
the configuration $\y \in \W_N^{h}$ at time $t=1$;
$\widetilde{\R}^{(N)}(1)=\y$.
Let $\widetilde{H}^{(N)}_N =
\max_{0 < t < 1} \widetilde{R}^{(N)}_N(t)$.
Then we can say that
$\P( \widetilde{H}^{(N)}_N < h)
=q^{(N)}_h(1, \y|\x)/q^{(N)} (1, \y|\x)$.
By definition of the $N$-noncolliding
Bessel bridges, $\r^{(N)}(t), 0 \leq t \leq 1$,
given in Sec.II.A, we can conclude that
\begin{equation}
\P(H^{(N)}_N < h)
= \lim_{|\x| \to 0, |\y| \to 0}
\frac{q^{(N)}_h (1, \y|\x)}{q^{(N)} (1, \y|\x)}.
\label{eqn:PHNh}
\end{equation}
As shown in Appendix A, the method of
the Schur function expansion \cite{KT04}
gives the following asymptotics for
$q^{(N)}(1, \y|\x)$ and
$q^{(N)}_h(1, \y|\x)$ in 
$|\x| \to 0, |\y| \to 0$,
\begin{eqnarray}
\label{eqn:asymq1}
q^{(N)}(1, \y|\x) &=&
\left( \frac{2}{\pi} \right)^{N/2}
\prod_{j=1}^{N} \frac{1}{(2j-1)!}
h_N^{\rm C}(\x) h_N^{\rm C}(\y)
\times \Big\{ 1+{\cal O}(|\x|, |\y|) \Big\}, \\
q^{(N)}_h(1, \y|\x) &=&
\left( \frac{2}{h} \right)^{N}
\left( \frac{\pi}{h} \right)^{2N^2}
\left\{ \prod_{j=1}^{N} \frac{1}{(2j-1)!} \right\}^2
h_N^{\rm C}(\x) h_N^{\rm C}(\y)
\nonumber\\
\label{eqn:asymq2}
&& \times \sum_{1 \leq n_1 < n_2 < \cdots < n_N}
\exp \left\{-\frac{\pi^2}{2 h^2} |\n|^2 \right\}
\Big\{ h_N^{\rm C}(\n) \Big\}^2
\times \Big\{ 1+{\cal O}(|\x|, |\y|) \Big\},
\end{eqnarray}
where $h_{N}^{\rm C}$ was defined by (\ref{eqn:hNC}).
Then Eq.(\ref{eqn:PHNh}) gives the following result.

\noindent{\bf Lemma 1.} \,
For $N \geq 1, h > 0$,
\begin{equation}
\P(H^{(N)}_N < h)
= c_N h^{-N(2N+1)}
\sum_{1 \leq n_1 < n_2 < \cdots < n_N}
\Big\{ h_N^{\rm C}(\n) \Big\}^2
\exp \left\{-\frac{\pi^2}{2 h^2} |\n|^2 \right\},
\label{eqn:result1}
\end{equation}
where
$c_N=2^{N/2} \pi^{N(4N+1)/2}/ \prod_{j=1}^{N} (2j-1)!$.
\vskip 0.5cm
\noindent{\bf Remark 1.} \,
This expression is exactly the same as Eq.(5) in \cite{SMCRF08},
which was derived by the path-integral method using 
a Selberg's integral.
Here we would like to put emphasis on the resemblance
between the summand of (\ref{eqn:result1}) and
(\ref{eqn:pN2}).
As mentioned in Sec.II.B, Eq.(\ref{eqn:pN2})
is the same as the probability density of the
eigenvalue-distribution of random matrices
in the class C (with variance $t(1-t)$).
The exponent of the factor $h^{-N(2N+1)}$
in (\ref{eqn:result1}) is the dimension of the
space ${\cal H}^{\rm C}(2N)$.
Another evidence to show the hidden symmetry
of the present maximum-value problem is the following.
The character of the irreducible representation
corresponding to a partition $\mu$
of the symplectic Lie algebra is given by \cite{FH91}
\begin{equation}
{\rm sp}_{\mu}(\bm{x})
=\frac{\displaystyle{\det_{1 \leq j, k \leq N}
\Big[x_j^{\mu_k+N-k+1}-x_j^{-(\mu_k+N-k+1)} \Big]}}
{\displaystyle{\det_{1 \leq j, k \leq N}
\Big[x_j^{N-k+1}-x_j^{-(N-k+1)}\Big]}}.
\label{eqn:sp1}
\end{equation}
If we set $n_j=\mu_{N-j+1}+j, 1 \leq j \leq N$,
and $x_j=1, 1 \leq j \leq N$,
(\ref{eqn:sp1}) gives
(see, for example,
Eq. (3.33) in \cite{KT03} and
Eq. (3.10) in \cite{KGV03}),
$$
{\rm sp}_{\mu}(1,1, \cdots, 1)
= \frac{h_N^{\rm C}(\n)}
{\prod_{j=1}^{N}(2j)!}.
$$
The above observations imply that 
the maximum-value problems of noncolliding diffusion
problems will be related with some enumerative 
problems of combinatorics in the ensembles
of irreducible representations of symmetry,
which can be regarded as discrete version of
random matrix ensembles.
The wall restriction for paths
({\it i.e.}, the stay-positive condition
for particles) is mapped into
the symplectic (the class C) structure
in the present case.
As pointed out in \cite{SMCRF08},
the trivial fact 
$\lim_{h \to \infty} \P(H^{(N)}_N < h)=1$
for (\ref{eqn:result1}) gives a version
of Selberg integral \cite{Sel44,Mac82},
\begin{eqnarray}
&&\lim_{\delta \to 0}
\sum_{n_1=-\infty}^{\infty} \cdots
\sum_{n_N=-\infty}^{\infty} 
\prod_{1 \leq j < k \leq N}
\Big\{(\delta n_j)^2-(\delta n_k)^2 \Big\}^2
\prod_{\ell=1}^{N} \Big\{ 
(\delta n_{\ell})^2 e^{-(\delta n_{\ell})^2/2}
\delta \Big\}
\nonumber\\
&=& \int_{-\infty}^{\infty} \cdots \int_{-\infty}^{\infty} 
\prod_{1 \leq j < k \leq N}
(x_j^2-x_k^2)^2 
\prod_{\ell=1}^{N} \Big\{
x_{\ell}^2 e^{-x_{\ell}^2/2} d x_{\ell} \Big\}
\nonumber\\
&=& (2 \pi)^{N/2} N ! \prod_{j=1}^{N} (2j-1)!,
\label{eqn:Selberg}
\end{eqnarray}
which is the special case with $\gamma=1$ and
$\alpha=3/2$ of Eq.(17.6.6) in \cite{Meh04}.
\vskip 0.5cm

\subsection{Determinantal expressions and
Jacobi's theta function}

From (\ref{eqn:result1}), we can obtain the
following determinantal expression
for the distribution function of $H^{(N)}_N$.

\noindent{\bf Proposition 2.} \,
For $N \geq 1, h > 0$,
\begin{eqnarray}
\P(H^{(N)}_N < h) &=& 
c_N h^{-N(2N+1)} 
\det_{1 \leq j, k \leq N}
\left[ \sum_{n=1}^{\infty} n^{2(j+k-1)}
e^{-\pi^2 n^2 /(2h^2)} \right]
\nonumber\\
&=& \frac{(-1)^{N} 2^{-N/2} \pi^{N(2N+1)/2}}{\prod_{j=1}^{N} (2j-1)!}
h^{-N(2N+1)} \tau_{N} 
\left( \frac{\pi}{2 h^2} \right),
\label{eqn:result2}
\end{eqnarray}
where
$$
\tau_{N}(u)=\det_{1 \leq j, k \leq N}
\left[ \frac{\partial^{j+k-1}}{\partial u^{j+k-1}}
\theta(u) \right] \quad \mbox{with} \quad
\theta(u)=\sum_{n=-\infty}^{\infty} e^{-\pi n^2 u}.
$$
\vskip 0.5cm
\noindent The proof is given in Appendix B.

Now we consider a version of Jacobi's theta function
\begin{eqnarray}
\vartheta_3(x,y) &=& \sum_{n=-\infty}^{\infty}
q^{n^2} z^{2n}
\nonumber\\
&=& \sum_{n=-\infty}^{\infty} e^{2 \pi i x n+\pi i y n^2},
\quad {\rm Im} \, y > 0,
\label{eqn:theta3}
\end{eqnarray}
where we have set
$z=e^{x \pi i}$ and $q=e^{y \pi i}$.
The following functional equation is satisfied
(see Sec.10.12 in \cite{AAR99}, Sec.A.3.1 in \cite{Ful07}),
\begin{equation}
\vartheta_3(x,y)
= \vartheta_3(x/y, -1/y) e^{-\pi i x^2/y}
\sqrt{\frac{i}{y}}.
\label{eqn:theta3b}
\end{equation}
From this equation, we will obtain the
following equalities.

\noindent{\bf Lemma 3.} \,
For $\alpha=0,1,2,\cdots $,
\begin{equation}
\sum_{n=-\infty}^{\infty} n^{\alpha} 
e^{-\pi n^2/\eta^2 + 2 \pi i \xi n/\eta^2} =
 \frac{i^{\alpha} \eta^{\alpha +1}}
 {2^{\alpha} \pi^{\alpha/2}}
 \sum_{n=-\infty}^{\infty} 
 H_{\alpha} \Big(\sqrt{\pi} n \eta+ \sqrt{\pi} \xi/ \eta \Big) 
  e^{-(\sqrt{\pi} n \eta + \sqrt{\pi} \xi /\eta)^2},
\label{eqn:theta3c}
\end{equation}
where $H_{\alpha}(x)$ is the $\alpha$-th Hermite
polynomial defined by (\ref{eqn:Hermite1}).

\noindent{\it Proof.} \quad
Setting $x=i \xi+ \eta^2 u/(2 \pi), y=i \eta^2$,
$\xi, \eta \in \R$,
in (\ref{eqn:theta3b}) gives
$$
\vartheta_3 \left(i \xi+\frac{\eta^2 u}{2\pi},i\eta^2\right) 
=
\vartheta_3 \left( \frac{\xi}{\eta^2}
-\frac{iu}{2\pi},\frac{i}{\eta^2} \right)
e^{\pi ( \xi/\eta- i \eta u/(2\pi) )^2} \frac{1}{\eta} 
$$
By definition (\ref{eqn:theta3}), this implies
\begin{equation}
\sum_{n=-\infty}^{\infty}  
e^{- \pi n^2 / \eta^2 +(2\pi i \xi / \eta^2+u)n}
= \eta \sum_{n=-\infty}^{\infty} 
 e^{-( \sqrt{\pi} \eta n +\sqrt{\pi}\xi/ \eta
 - i \eta u/(2 \sqrt{\pi}))^2}  
\label{eqn:eqA}
\end{equation}
We note that the generating function of Hermite polynomial
is given as
$$
e^{2 u x- u^2} = \sum_{n=0}^{\infty}H_n(x)\frac{u^n}{n!},
$$
Then (\ref{eqn:eqA}) becomes
\begin{eqnarray}
&&\sum_{\alpha=0}^{\infty}
\sum_{n=-\infty}^{\infty} 
\frac{(n u)^{\alpha}}{\alpha !}
e^{- \pi n^2/ \eta^2 + 2 \pi i \xi n / \eta^2}
\nonumber\\
&& \qquad
= \eta \sum_{\alpha=0}^{\infty}
\sum_{n=-\infty}^{\infty} 
 H_{\alpha} \Big(\sqrt{\pi} n \eta+ \sqrt{\pi} \xi/\eta \Big) 
 \left(\frac{i \eta u}{2 \sqrt{\pi}} \right)^{\alpha}
 \frac{1}{\alpha !}
  e^{-(\sqrt{\pi} n \eta +\sqrt{\pi} \xi / \eta)^2}.
\nonumber
\end{eqnarray}
Since this equality holds for any value of $u$,
Lemma 3 is proved.

By setting $\xi=0, \eta=h \sqrt{2/\pi}$ and
$\alpha= 2 \ell$ in Eq.(\ref{eqn:theta3c})
of Lemma 3, we have the equalities
\begin{equation}
\sum_{n=1}^{\infty} n^{2 \ell}
e^{-\pi^2 n^2/(2h^2)}
= \frac{(-1)^{\ell}}{2^{\ell+1/2} \pi^{2 \ell+1/2}}
h^{2 \ell+1}
\sum_{n=-\infty}^{\infty} H_{2 \ell}
(\sqrt{2} n h) e^{-2 n^2 h^2}
\label{eqn:eqB}
\end{equation}
for $\ell=0,1,2, \cdots$.
Then Eq.(\ref{eqn:result2}) is transformed into
the following other
determinantal expression, which was
announced in Introduction.

\noindent{\bf Theorem 4.} \,
For $N \geq 1, h > 0$, 
\begin{equation}
\P(H^{(N)}_N < h) 
= \frac{(-1)^N}{2^{N^2} \prod_{j=1}^{N} (2j-1)!}
\det_{1 \leq j, k \leq N}
\left[ \sum_{n=-\infty}^{\infty} H_{2(j+k-1)}
(\sqrt{2} n h) e^{-2 n^2 h^2} \right].
\label{eqn:Main2}
\end{equation}
\vskip 0.5cm

As special cases of (\ref{eqn:Main2})
with $N=1$ and $N=2$, we have
$$
\P(H^{(1)}_1 < h) = -\frac{1}{2} 
\sum_{n=-\infty}^{\infty}
H_2(\sqrt{2} n h) e^{-2 h^2 n^2}
$$
and
$$
\P(H^{(2)}_2 < h) = \frac{1}{2^{4} \times 3!}
\sum_{n_1=-\infty}^{\infty} \sum_{n_2=-\infty}^{\infty}
e^{-2 h^2 (n_1^2+n_2^2)}
\det \left[ 
\begin{array}{cc}
H_2(\sqrt{2} n_1 h) & H_4(\sqrt{2} n_1 h) \cr
H_4(\sqrt{2} n_2 h) & H_6(\sqrt{2} n_2 h)
\end{array}
\right].
$$
Since 
$H_2(x) = 4x^2 -2, 
H_4(x) = 16 x^4-48 x^2+12$ and
$H_6(x) = 64 x^6 -480 x^4 + 720 x^2 -120$,
they give
\begin{equation}
\P(H^{(1)}_1 < h ) = \sum_{n=-\infty}^{\infty}
(1-4 h^2 n^2) e^{-2 h^2 n^2}
\label{eqn:H11}
\end{equation}
and
\begin{eqnarray}
\P(H^{(2)}_2 < h) &=& \sum_{n_1=-\infty}^{\infty}
\sum_{n_2=-\infty}^{\infty}
e^{-2h^2(n_1^2+n_2^2)}
\left\{ 1-16h^2 n_1^2+24 h^4 n_1^4 + 24 h^4 n_1^2 n_2^2
-\frac{32}{3} h^6 n_1^6 \right.
\nonumber\\
&& \qquad
\left. - 32 h^6 n_1^4 n_2^2 + \frac{128}{3} h^8 n_1^6 n_2^2
-\frac{128}{3} h^8 n_1^4 n_2^4 \right\}.
\label{eqn:H22}
\end{eqnarray}
From (\ref{eqn:H11}) we will obtain (\ref{eqn:pH1}).
Eq.(\ref{eqn:H22}) is exactly the same as the result reported
as Lemma 3.1 in our previous paper \cite{KIK08}.

\noindent{\bf Remark 2.} \,
Since the derivative of the distribution function
$\P(H^{(N)}_N < h)$ with respect to $h$ gives
the probability density for $H^{(N)}_N \in dh$,
the $s$-th moment of $H^{(N)}_N, s=1,2, \dots$,
is calculated as
\begin{eqnarray}
&& \langle (H^{(N)}_N)^s \rangle
= \int_{0}^{\infty} h^{s} \left(
\frac{d}{dh} \P(H^{(N)}_N < h) \right) dh
\nonumber\\
&& \quad 
= s \int_{0}^{\infty} h^{s-1}
\Big\{ 1- \P(H^{(N)}_N < h) \Big\} dh,
\label{eqn:momentA}
\end{eqnarray}
where the integral by part was done.
If we insert the expression (\ref{eqn:result2})
into (\ref{eqn:momentA}), we have a determinantal expression
for the $s$-th moment. It is essentially the same as the
expression of Feierl given to the dominant
term in the long-step asymptotics of the moment
of the height distribution of watermelons
with a wall.
See the function $\kappa^{(p)}_s$ given in
Theorem 1 of \cite{Fei08a}.
Moreover, the determinantal expression for the
central limit theorem of the height distribution
of the watermelons with a wall given by Feierl
(Eq.(28) in Theorem 2) \cite{Fei08a} can be
identified with our second determinantal
expression (\ref{eqn:Main2}),
since the functions $\phi_k(z)$ used there are
nothing but the Hermite polynomials.
We have enjoyed the perfect coincidence
of the results obtained by the two different
routes to the problem.
See \cite{KIK08} for more detailed
discussion on the relationship
between our treatment and that by
Fulmek \cite{Ful07} and Feierl \cite{Fei07,Fei08a,Fei08b}.

\section{Numerical Study}
\subsection{Bessel bridge realized by eigenvalue process}

We have prepared a computer program to generate 
samples of paths of $b(t), 0 \leq t \leq 1$;
the Brownian bridge with duration 1 starting from the origin,
in which each sample path is approximated by 
random walk with 10000 steps.

\begin{figure}
\includegraphics[width=0.6\linewidth]{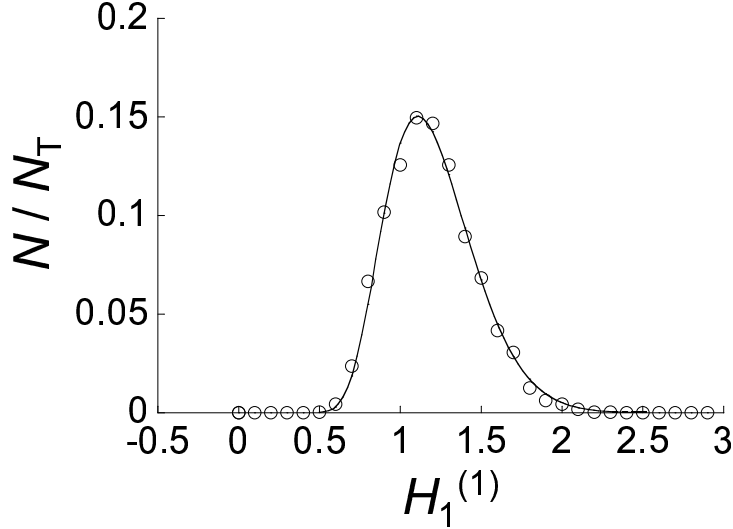}
\caption{Comparison of the distribution of the maximum values
$H^{(1)}_1$ of the 3-dimensional Bessel bridge
between the numerical simulation (plotted by circles)
and the theoretical values (given by curve).}
\label{fig:Fig4}
\end{figure}

Assume that we have generated three independent
Brownian bridges, $b_j(t), 0 \leq t \leq 1,
j=1,2,3$, by this computer program.
Then we consider the following $2 \times 2$ matrix-valued
process,
\begin{equation}
C^{(1)}(t) = \left(
\begin{array}{cc}
b_{1}(t) & b_{2}(t) + i b_{3}(t) \\
b_{2} - i b_{3}(t) & - b_{1}(t)
\end{array}
\right).
\label{eqn:C1t}
\end{equation}
It is easy to see that the eigenvalue process
of $C^{(1)}(t)$ is given by the
positive and negative pair of eigenvalues
$\vlambda^{(2)}(t)=(\lambda^{(1)}_1(t), -\lambda^{(1)}_1(t))$
with
\begin{equation}
\lambda^{(1)}_1(t)=
\sqrt{ (b_1(t))^2+(b_2(t))^2+(b_3(t))^2}.
\label{eqn:lamb11}
\end{equation}
The positive eigenvalue-process (\ref{eqn:lamb11})
gives the numerical realization of
$r(t), 0 \leq t \leq 1$;
the 3-dimensional Bessel bridge.
Actually the sample of path of $r(t)$ shown in
Fig.\ref{fig:Fig1} given in Introduction was numerically
drawn by this method.
In order to check the validity of the present
numerical method to simulate the Bessel bridge,
we have generated 1000 samples of paths and studied
the distribution of their maximum values
$H^{(1)}_1= \max_{0 < t < 1} r(t)$ numerically.
The obtained result is plotted in Fig.\ref{fig:Fig4}.
There the exact curve obtained from
Eq.(\ref{eqn:pH1}) is also shown.
The coincidence is excellent.

\subsection{Means and variances of $H^{(N)}_N$'s}

\begin{table}[t]
\begin{center}
\begin{tabular}{c p{0.5cm} c p{0.5cm}c p{0.5cm}c p{0.5cm}c p{0.5cm}c p{0.5cm}c p{0.5cm}c p{0.5cm}c p{0.5cm}c p{0.5cm}c p{0.5cm}c p{0.5cm}c}
\cline{1-13}
$N$ & & $m^{(N)}_N$ & & $v^{(N)}_N$ 
& & F $m^{(N)}_N$ & & F $v^{(N)}_N$ 
& & KIK $m^{(N)}_N$ & & KIK $v^{(N)}_N$\\
\hline\hline
1 & & 1.251 & & 0.0774 & & 1.2533 & & 0.0737 & & 1.253314 & & 0.074138  \\

2 & & 1.819 & & 0.0732 & & 1.8222 & & 0.0746 & & 1.822625 & & 0.073194 \\

3 & & 2.262 & & 0.0704 & & 2.2677 & & 0.0720 & &          & & \\

4 & & 2.641 & & 0.0664 & & 2.6460 & & 0.0692 & &          & & \\

5 & & 2.979 & & 0.0640 & & 2.9805 & & 0.0656 & &          & & \\

6 & & 3.280 & & 0.0624 & &        & &        & &          & & \\

7 & & 3.558 & & 0.0600 & &        & &        & &          & & \\

8 & & 3.817 & & 0.0556 & &        & &        & &          & & \\

9 & & 4.057 & & 0.0570 & &        & &        & &          & & \\

10 & & 4.291 & & 0.0543 & &        & &        & &          & & \\

20 & & 6.146 & & 0.0409 & &        & &        & &          & & \\

30 & & 7.597 & & 0.0396 & &        & &        & &          & & \\

40 & & 8.790 & & 0.0384 & &        & &        & &          & & \\

50 & & 9.841 & & 0.0364 & &        & &        & &          & & \\

60 & & 10.806 & & 0.0354 & &        & &        & &          & & \\

70 & & 11.678 & & 0.0310 & &        & &        & &          & & \\
\cline{1-13}
\end{tabular}
\caption{Comparison of the values of $m_{N}^{(N)}$ 
and $v_{N}^{(N)}$ obtained by the present numerical method
and the exact values. The values with `F' are read from Table 1 in 
\cite{Fei08a} and
those with `KIK' are from Table 1 in \cite{KIK08}.}
\label{tab:s=1_2}
\end{center}
\end{table}

Note that the $2 \times 2$ matrix (\ref{eqn:C1t}) can be
considered as the special case of (\ref{eqn:Ct1})
with $N=1$.
If we prepare ten independent Brownian bridges,
$\{b_j(t)\}_{j=1}^{10}$, numerically,
we can simulate the $4 \times 4$ matrix-valued
Brownian bridge in ${\cal H}^{\rm C}(4)$,
$$
C^{(2)}(t) = \left(
\begin{array}{cccc}
b_{1}(t) & \frac{1}{\sqrt{2}} (b_{2}(t) + i b_{3}(t)) 
& b_{5}(t) + i b_{6}(t) 
& \frac{1}{\sqrt{2}} (b_{7}(t) + i b_{8}(t)) \\
\frac{1}{\sqrt{2}} (b_{2}(t) - i b_{3}(t)) & b_{4}(t) 
& \frac{1}{\sqrt{2}} (b_{7}(t) + i b_{8}(t)) & b_{9}(t) + i b_{10}(t) \\
b_{5}(t) - i b_{6}(t) & \frac{1}{\sqrt{2}} (b_{7}(t) - i b_{8}(t))  
& - b_{1} & -\frac{1}{\sqrt{2}} (b_{2}(t) - i b_{3}(t))  \\
\frac{1}{\sqrt{2}} (b_{7}(t) - i b_{8}(t)) & b_{9}(t) - i b_{10}(t) 
& -\frac{1}{\sqrt{2}} (b_{2}(t) + i b_{3}(t)) & - b_{4}(t)
\end{array}
\right),
$$
which is the $N=2$ case of (\ref{eqn:Ct1}).
By tracing the two positive eigenvalues
$\vlambda^{(2)}_{+}(t)=(\lambda^{(2)}_1(t), \lambda^{(2)}_2(t)),
0 < \lambda^{(2)}_1(t) < \lambda^{(2)}_2(t), 
0 \leq t \leq 1$,
we can simulate the noncolliding paths of two 
Bessel bridges,
$\r^{(2)}(t)=(r^{(2)}_1(t), r^{(2)}_2(t))$ and
statistical data of $H^{(2)}_{2}=\max_{0 < t < 1} r^{(2)}_2(t)$
can be obtained.
In general, we can simulate the $N$-noncolliding Bessel bridges
$\r^{(N)}(t), 0 \leq t \leq 1$, by using numerical data of
independently generated $N(2N+1)$ Brownian bridges
$\{b_j(t)\}_{j=1}^{N(2N+1)}$
and by tracing the $N$ positive eigenvalues of the
$2N \times 2N$ matrix.

From now on, we use the notations
\begin{eqnarray}
m^{(N)}_k &=& \langle H^{(N)}_k \rangle,
\nonumber\\
v^{(N)}_k &=& {\rm var} ( H^{(N)}_k )
= \langle (
H^{(N)}_k-m^{(N)}_k )^2 \rangle,
\quad 1 \leq k \leq N
\label{eqn:mv}
\end{eqnarray}
for the means and variances of maximum values
of paths.
Table I shows the numerical results for the outermost
paths $k=N$
up to $N=70$, where averages have been calculated over
1000 samples.
The present numerical results are consistent with the
exact values, which can be read from the
previous papers by Feierl \cite{Fei08a}
and by the present authors \cite{KIK08}.

\subsection{$N \to \infty$ asymptotics of the outermost paths}

\begin{figure}
\includegraphics[width=0.6\linewidth]{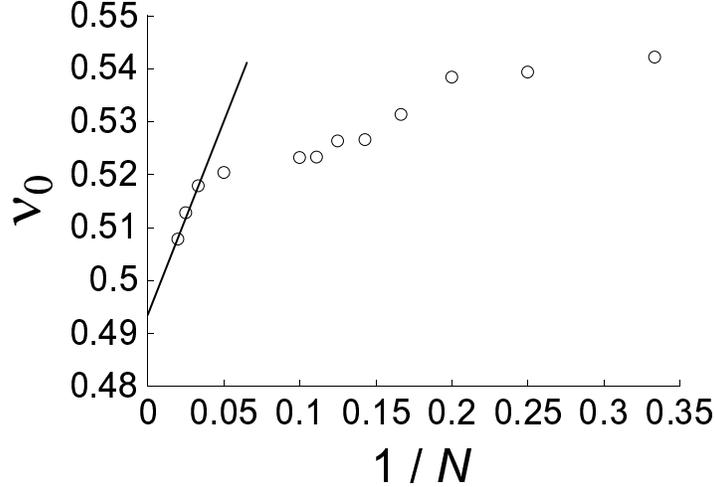}
\caption{$1/N$-plot of the estimated values of $\nu_0$}
\label{fig:Fig5}
\end{figure}
\begin{figure}
\includegraphics[width=0.6\linewidth]{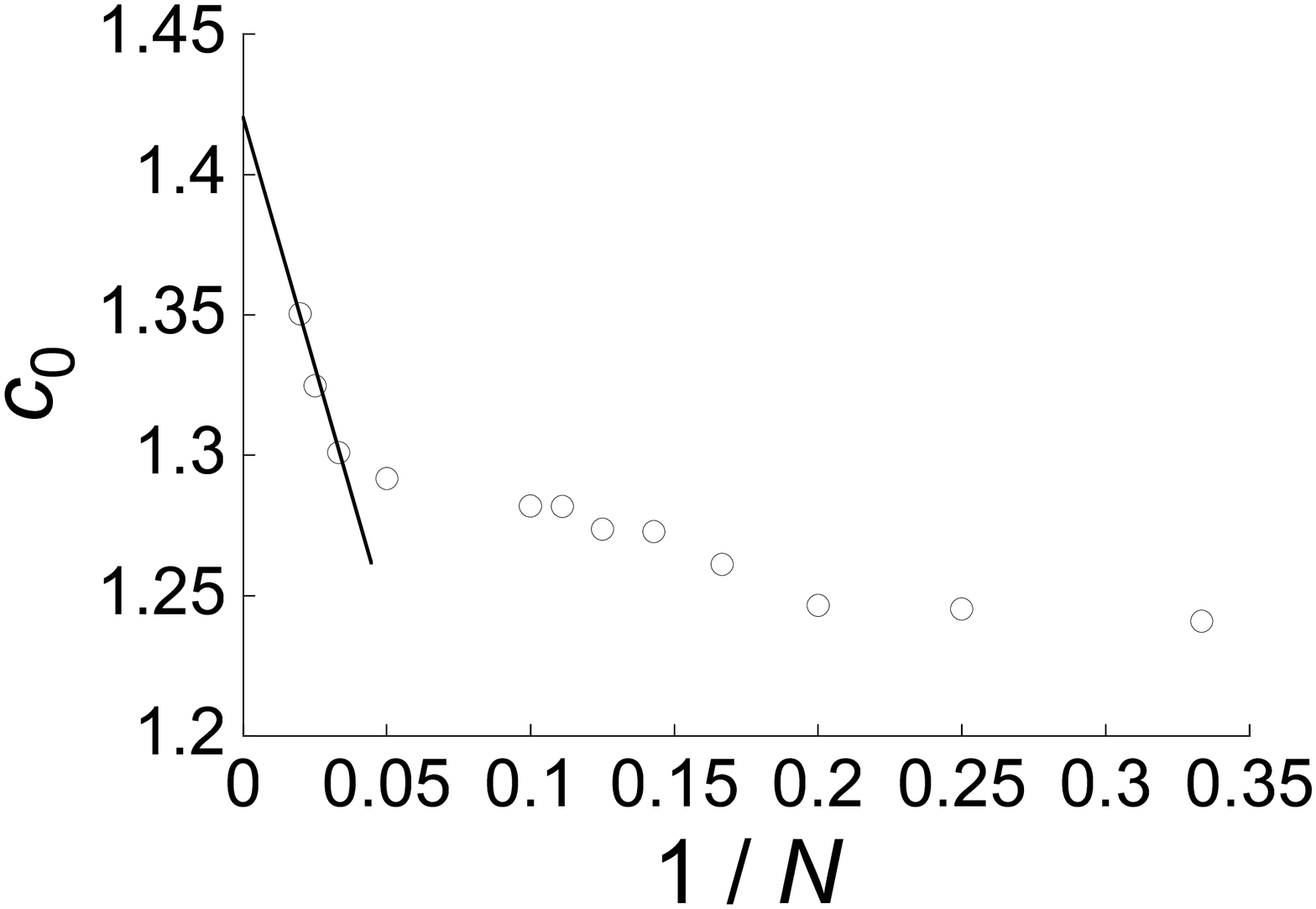}
\caption{$1/N$-plot of the estimated values of $c_0$ }
\label{fig:Fig6}
\end{figure}
\begin{figure}
\includegraphics[width=0.6\linewidth]{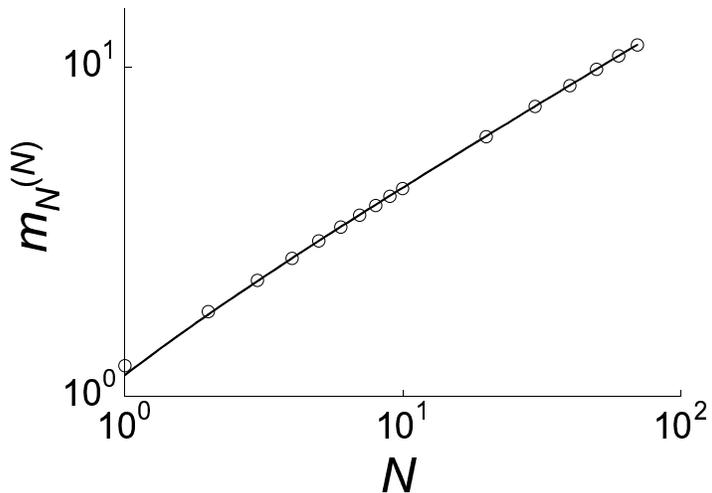}
\caption{Log-log plot of $m^{(N)}_N$ vs. $N$ for $N$-noncolliding
Bessel bridges. }
\label{fig:Fig7}
\end{figure}

Now we study the $N \to \infty$ asymptotics of $m^{(N)}_N$.
We assume the form
\begin{equation}
m^{(N)}_N= c_0 N^{\nu_0},
\quad N \gg 1.
\label{eqn:fitA1}
\end{equation}
We have prepared a set of successive five results 
of the computer simulation,
$(m^{(N_1)}_{N_1}, m^{(N_2)}_{N_2}, m^{(N_3)}_{N_3}, 
m^{(N_4)}_{N_4}, m^{(N_5)}_{N_5})$
and fitted them to the relation (\ref{eqn:fitA1})
to estimate the exponent $\nu_0$ and 
the coefficient $c_0$
by the least-square fitting.
We have observed that the estimated values
of $\nu_0$ and $c_0$ change rather systematically
as the values $N_j, 1 \leq j \leq 5$, increase.
Figures \ref{fig:Fig5} and \ref{fig:Fig6} show the 
dependence of the estimated values of $\nu_0$ and $c_0$
on $N=N_3$. In these $1/N$-plots \cite{Gut89} 
of the estimated values,
we made linear-fitting of the largest
three plots as shown by the lines in the figures
and obtained the values
$\nu_0 = 0.493$ and $c_0=1.42$.
They are consistent with Eq.(\ref{eqn:Schehr1});
$\nu_0=1/2$ and $c_0=\sqrt{2}=1.414 \dots$.
In Fig.\ref{fig:Fig6} we can see that the
plots in the intermediate region
$0.1 < 1/N < 0.2$ give the
values $1.25 < c_0 < 1.3$.
They may correspond to the value
$\sqrt{1.67} \simeq 1.29$ found in the
estimate (\ref{eqn:BM1}) by Bonichon and Mosbah \cite{BM03},
which was claimed by Schehr {\it et al.} as
the pre-asymptotic behavior \cite{SMCRF08}.
Figure 7 shows the log-log plot of $m^{(N)}_N$
versus $N$, where the curve is obtained by
fitting the function
\begin{equation}
m^{(N)}_N = \sqrt{2N}
+ c_1 N^{-\nu_1}
\quad \mbox{with} \quad
\nu_1=\frac{1}{6}
\label{eqn:nu1}
\end{equation}
to the data. The fitting parameter $c_1$ is determined
as $c_1 \simeq 0.253$.

\begin{figure}
\includegraphics[width=0.6\linewidth]{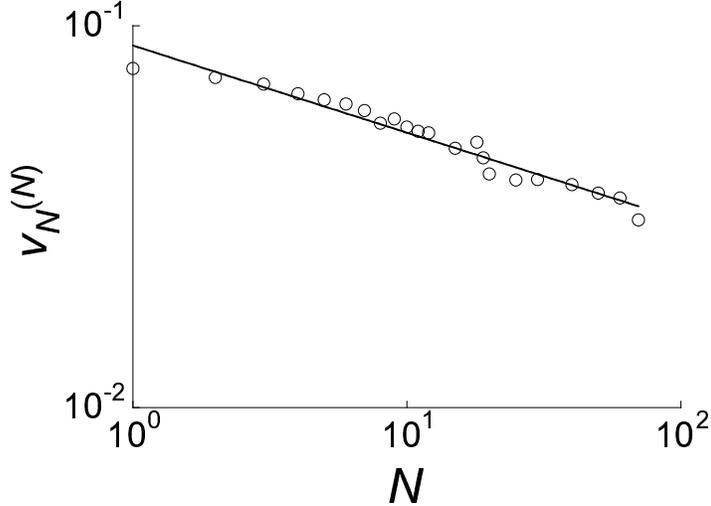}
\caption{Log-log plot of $v^{(N)}_N$ vs. $N$ for $N$-noncolliding
Bessel bridges.}
\label{fig:Fig8}
\end{figure}

We have also studied the asymptotics of variance
$v^{(N)}_N$ in $N \to \infty$.
Figure \ref{fig:Fig8} shows the log-log plot.
By the least-squares fitting, we have obtained
\begin{equation}
v^{(N)}_N \simeq 0.09 N^{-0.23}.
\label{eqn:vN_fit}
\end{equation}
The negative exponent implies
$v^{(N)}_N \to 0$ in $N \to \infty$.

\subsection{On inner paths }

Using the numerical data of paths
of $N$-noncolliding Bessel bridges
generated by the method mentioned in Sec.IV-B,
we can examine the statistics of the maximum values
attained in the time interval $(0,1)$ of not only
the outermost paths but also all inner paths.
For example, the means $m^{(N)}_k$
and the variances $v^{(N)}_k$ for all paths
$1 \leq k \leq N$ are given in Table II for
$N=10$.

\begin{table}[h]
\begin{center}
\begin{tabular}{c p{0.5cm} c p{0.5cm} c}
\cline{1-5}
$k$  &  & $m_{k}^{(10)}$ &  & $v_{k}^{(10)}$ \\
\hline\hline
1 &  & 0.547 &  & 0.00528 \\
2 &  & 0.891 &  & 0.00778 \\
3 &  & 1.240 &  & 0.00971 \\
4 &  & 1.598 &  & 0.0126 \\
5 &  & 1.958 &  &  0.0147 \\
6 &  & 2.337 &  &  0.0170 \\
7 &  &  2.735 &  &  0.0209 \\
8 &  &  3.168 &  & 0.0262 \\
9 &  & 3.659 &  & 0.0320 \\
10 &  & 4.291 &  &  0.0553 \\
\cline{1-5}
\end{tabular}
\caption{Numerical values of $m_{k}^{(10)}$ and $v_{k}^{(10)}$
evaluated by computer simulations}
\end{center}
\end{table}

\begin{figure}
\includegraphics[width=0.6\linewidth]{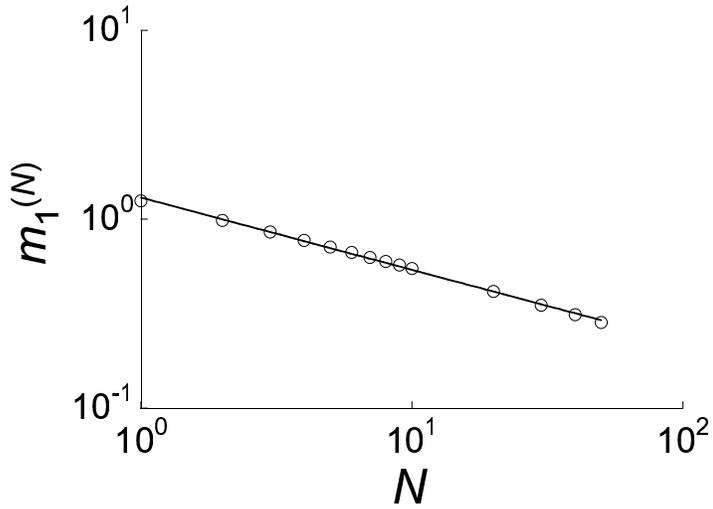}
\caption{Log-log plot of $m^{(N)}_1$ vs. $N$ for 
$N$-noncolliding Bessel bridges.}
\label{fig:Fig9}
\end{figure}

\begin{figure}
\includegraphics[width=0.6\linewidth]{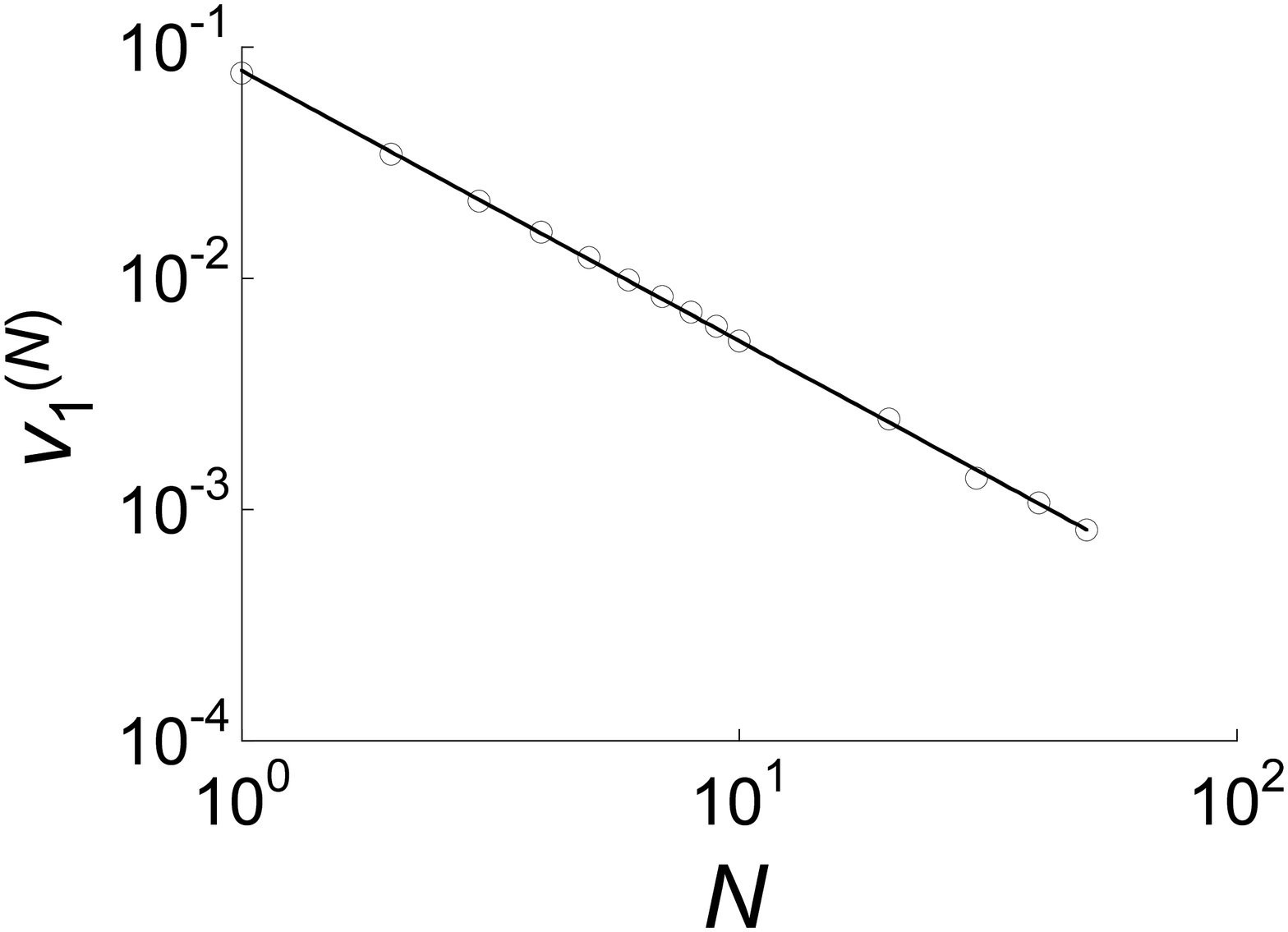}
\caption{Log-log plot of $v^{(N)}_1$ vs. $N$ for 
$N$-noncolliding Bessel bridges.}
\label{fig:Fig10}
\end{figure}

Now we report our observation of the $N$-dependence 
for the maximum values of inner paths.
Figures \ref{fig:Fig9} and \ref{fig:Fig10} show the $N$-dependence
of the means and variances for the
maximum values of the innermost path,
$H^{(N)}_1$.
By the log-log plots, we have obtained the following
power-law behavior of the $N$-dependence.
\begin{eqnarray}
\label{eqn:fit2}
m_{1}^{(N)} \sim N^{- 0.38}, \quad
v_{1}^{(N)} \sim N^{- 1.17}.
\end{eqnarray}


\begin{figure}
\includegraphics[width=0.6\linewidth]{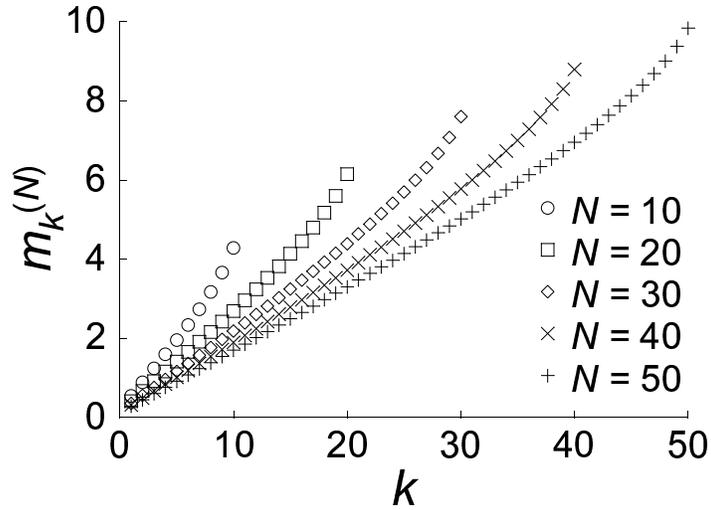}
\caption{Dependence of $m_{k}^{(N)}$ on $k$
shown for various $N$.}
\label{fig:Fig11}
\end{figure}
\begin{figure}
\includegraphics[width=0.6\linewidth]{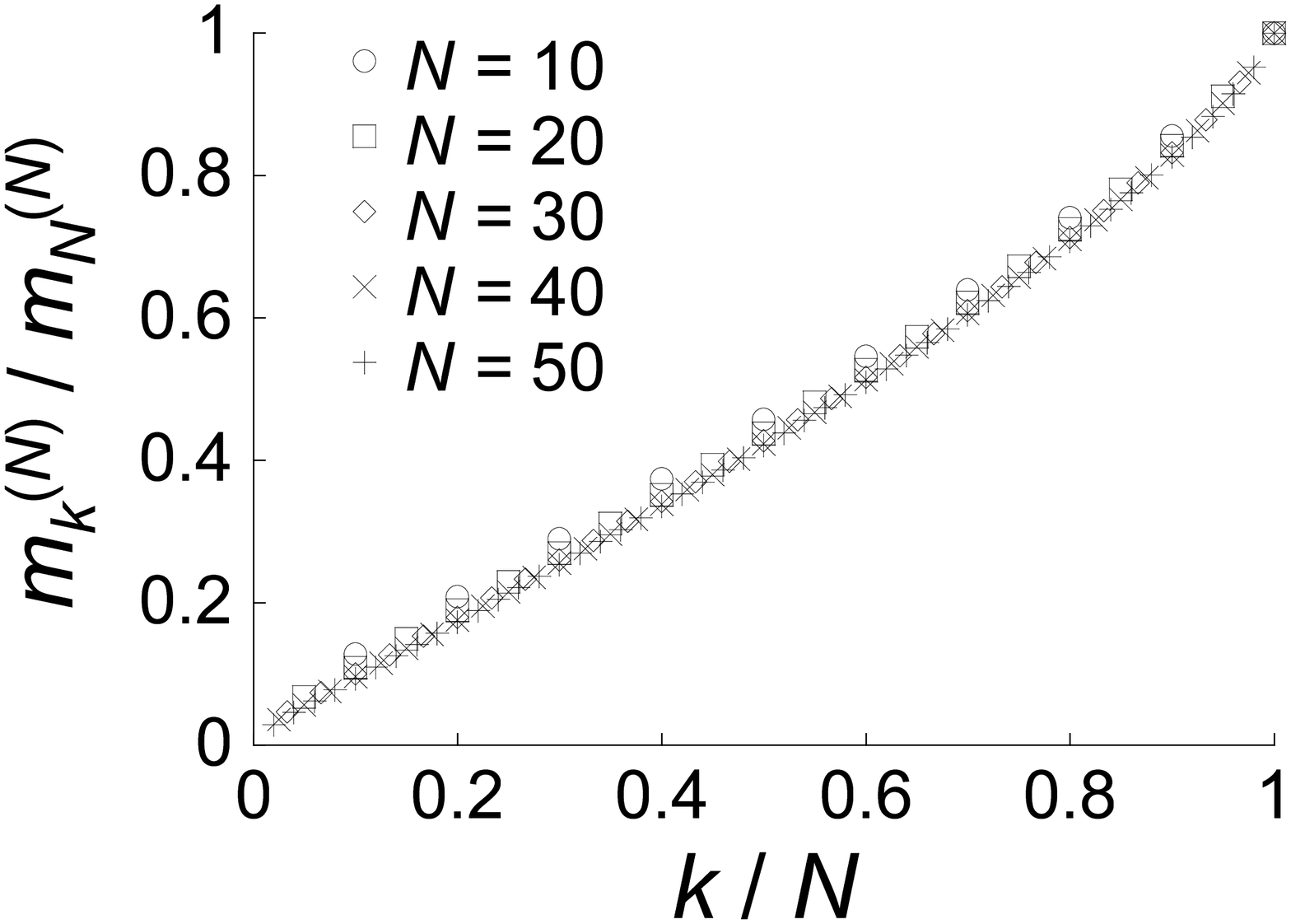}
\caption{Data collapse for the maximum values of inner
particles for $N$-noncolliding Bessel bridges.}
\label{fig:Fig12}
\end{figure}

Figure \ref{fig:Fig11} shows the dependence of $m_{k}^{(N)}$ on
$k$, which is the index of paths counting from 
inner to outer.
We can see that all plotted curves in Fig. \ref{fig:Fig11}
have a common feature as a function of $k$.
This fact is clarified by Fig.\ref{fig:Fig12}, 
in which the data collapse
\cite{KSOYHM05} is shown by plotting
the quantity $m_{k}^{(N)} / m_{N}^{(N)}$ against $k / N$.
The result implies  that
there is a universal function $f(x)$ such that
the following relation holds for sufficiently large $N$,
\begin{align}
\frac{m_{k}^{(N)}}{m_{N}^{(N)}} \sim f\left( \frac{k}{N} \right).
\label{eqn:scalingf}
\end{align}
The scaling function $f(x)$ has two regions separated by a crossover
point $x^{\ast}$.
For small $x \ll x^{\ast}$, the scaling function $f(x)$ 
behaves as a linear function.
On the other hand, for large $x \gg x^{\ast}$, $f(x)$ does not 
behave as a simple linear function.
We have estimated the following by numerical fitting.
\begin{align}
f(x) &\sim 
\begin{cases}
x & \quad (x \ll x^{\ast}),\\
a^{x} & \quad (x \gg x^{\ast}) \quad
\mbox{with} \quad a \simeq 5.1.
\end{cases}
\label{eqn:scaling_m}
\end{align}

\section{Concluding Remarks}

In this paper we have reported the exact and the numerical
results on the maximum-value distributions of paths
in the $N$-noncolliding Bessel bridges.
We have shown that the present maximum-value problem
for a version of vicious walk model
of statistical physics is related with
the random matrix theory, the representation
theory of symmetry, and the number theory.
There are a lot of open problems.
We will list up here some of them.

In the expression (\ref{eqn:phabs}) of the
transition probability density $p^{h}_{\rm abs}$
of the absorbing BM in an interval $(0,h)$,
a variable $n \in \N=\{1,2,3, \dots\}$ was 
introduced to indicate modes in the Fourier expansion.
When we consider the $N$-path systems,
a set of $N$ discrete variables
$\n=(n_1, n_2, \dots, n_N) \in \N^{N}$ is introduced.
Though the variables $\n$ are auxiliary,
since the physical quantities are given by
the summations over $\n$'s as shown in
(\ref{eqn:phabs}), (\ref{eqn:result1}),
(\ref{eqn:result2}) and (\ref{eqn:Main2}),
we have seen in the derivation of Lemma 1
given in Appendix A the discrete variables
$\n$ behave as duals of the continuous variables
$\x=(x_1, x_2, \dots, x_N)$ and
$\y=(y_1, y_2, \dots, y_N)$,
which are physical variables indicating the positions
of particles.
The correspondence between
the probability density of paths
(\ref{eqn:pN2}) given in the form of
that of eigenvalues of random matrices
in class C
and the distribution function of the
maximum value (\ref{eqn:result1})
given in the form of the
``partition function" of discrete variables
implies some duality relation.
The maximum- and minimum-value problems
of watermelons {\it without wall}
recently studied by Feierl \cite{Fei08b}
and by Schehr {\it et al.}\cite{SMCRF08}
are very interesting.
Systematic study will be desired
to clarify the duality between the noncolliding
path problems \cite{KT07b}
and their extreme-value problems
not only for bridges ({\it i.e.},
excursions, watermelon configurations)
\cite{TW07,Kuij08}
but also for meanders ({\it i.e.} 
star configurations) \cite{KT07a}.

As shown in Introduction, the distribution of
the maximum value of the Bessel bridge
$H^{(1)}_1=\max_{0 < t < 1} r(t)$,
and that of the position of Bessel bridge $r(1/2)$
at time $t=1/2$ are
quite different from each other, because of large
fluctuation of path.
For example, the mean value of the maximum
$\langle H^{(1)}_1 \rangle = \sqrt{\pi/2}
\simeq 1.2533$ is much bigger than
$\langle r(1/2) \rangle=\sqrt{2/\pi}
\simeq 0.7979$.
As shown by Fig.\ref{fig:Fig8}, however, the variance
$v^{(N)}_N$ will vanish in $N \to \infty$
as Eq.(\ref{eqn:vN_fit}).
Then we expect
$$
\langle H^{(N)}_N \rangle
\simeq \langle r^{(N)}_N(1/2) \rangle
\quad \mbox{in} \quad
N \to \infty.
$$
It is known that
$$
\langle r^{(N)}_N(t) \rangle
\simeq 2 \sqrt{2N t(1-t)},
\quad 0 \leq t \leq 1
\quad \mbox{in} \quad N \to \infty
$$
(see, for example, Eq.(2.26) in \cite{Kuij08},
in which we should put $a=0$ and use the relation
$\sqrt{x}=\lim_{N \to \infty}
\langle r^{(N)}_N(t) \rangle / \sqrt{2N}$.)
Then Schehr {\it et al.} concluded
Eq.(\ref{eqn:Schehr1}).
By taking the proper scaling limit associated with
$N \to \infty$, the fluctuation of the outermost
path $r^{(N)}_N(t), 0 \leq t \leq 1$,
defines the Airy process \cite{PS02,Joh03,TW03,TW07},
which obeys the Tracy-Widom distribution \cite{TW94}
at each time $0 < t < 1$.
The present study suggests us to study the maximum
value distribution of the Airy process.

We have used the exact expressions
(\ref{eqn:result1}), (\ref{eqn:result2}) and (\ref{eqn:Main2}) for distributions
of $H_N^{(N)}$ given in Sec.III
to check the validity of numerical 
calculations by our computer programs.
As mentioned in Remark 1, the expression 
(\ref{eqn:result1}) seems to be a ``partition function"
of some discrete statistical-mechanical
model with the Boltzmann weight
$\exp(-\pi^2 |{\bf n}|^2/(2h^2))$.
The expression (\ref{eqn:result2}) reminds us 
the bidirectional Wronskian solutions
of nonlinear equations.
And as demonstrated below Theorem 4,
the expression (\ref{eqn:Main2}) is useful to reproduce
the previous results reported in 
\cite{BPY01, Ful07, KIK08} for small $N$.
Deeper understanding of these expressions
in physics is desired.
At the present stage it is not obvious
how to discuss the asymptotics of these
expressions in $N \to \infty$.
We hope that our determinantal expressions
(\ref{eqn:result2}) and (\ref{eqn:Main2}) will be useful, since 
determinantal formulas play important roles 
in the random matrix theory in analyzing
the large-matrix limit \cite{BS98}, when they are
equipped with proper mathematical tools,
{\it e.g.} orthogonal polynomials, 
Fredholm determinants and so on
\cite{Meh04, KT07b}.

It will be challenging open problem to analyze
the maximum-value distributions for
the inner paths in the present systems.

\begin{acknowledgments}
One of the authors (M.K.) would like to thank
Thomas Feierl,
Markus Fulmek,
Michael Drmota,
Christian Krattenthaler,
Hideki Tanemura,
and Tomohiro Sasamoto
for useful discussions
on the present problem.
A part of the present work was done
during the participation of M.K.
in the ESI program ``Combinatorics and Statistical Physics"
(March and May in 2008).
M.K. expresses his gratitude for 
hospitality of the Erwin Schr\"odinger Institute 
(ESI) in Vienna
and for well-organization of the program
by Michael Drmota and Christian Krattenthaler.
The authors thank Gr\'egory Schehr
for sending their paper before publication.
M.K. is supported in part by
the Grant-in-Aid for Scientific Research (C)
(No.17540363) of Japan Society for
the Promotion of Science.
\end{acknowledgments}

\appendix
\section{Derivation of the asymptotics
(\ref{eqn:asymq1}) and (\ref{eqn:asymq2})}

By the multilinearity of determinant
\begin{eqnarray}
q^{(N)}(1,\y|\x) 
&=& \det_{1 \le j, k \le N} \left[ \frac{1}{\sqrt{2\pi}}
\Big\{e^{-(y-x)^2 / 2} - e^{-(y + x)^2 / 2}\Big\}\right] \nonumber\\
&=& \frac{1}{(2\pi)^{N / 2}}e^{-(|\x|^2 + |\y|^2) / 2} 
\det_{1 \le j, k \le N}\left[e^{y_jx_k} - e^{-y_jx_k}\right].
\nonumber
\end{eqnarray}
Here
\begin{eqnarray}
& &\det_{1 \le j, k \le N}\left[e^{y_jx_k} - e^{-y_jx_k}\right] \nonumber\\
&=& 2^{N}\prod_{j = 1}^{N}(x_jy_j)
\sum_{0 \le m_1 < m_2 < \cdots < m_N}
\prod_{j=1}^{N}\frac{1}{(2m_j + 1)!}
\det_{1 \le j, k \le N}\left[y_{j}^{2m_k}\right]
\det_{1 \le j, k \le N}\left[x_{j}^{2m_k}\right].
\nonumber
\end{eqnarray}
Now we change the variables 
in summation from ${m_j}$ to ${\mu_j}$ by $\mu_j = 
m_{N-j+1} - N + j, 1 \le j \le N$,
and introduce the Schur function
\cite{FH91,Mac95,Ful97}
\begin{equation}
s_{\mu}(\x) 
= \frac{\displaystyle{\det_{1 \le j, k \le N}
\left[x_{j}^{\mu_k + N -k}\right]}}
{\displaystyle{\det_{1 \le j, k \le N}\left[x_{j}^{N-k}\right]}},
\label{eqn:Schur}
\end{equation}
where the denominator is the Vandermonde determinant
\begin{equation}
\det_{1 \le j, k \le N}\left[x_{j}^{N-k}\right] 
= \prod_{1 \le j, k \le N}(x_j - x_k).
\label{eqn:Vand}
\end{equation}
The Schur function expansion 
is readily performed as (see \cite{KT07b,KT04})
\begin{eqnarray}
q^{(N)}(1, \y|\x) &=& 
\Big(\frac{2}{\pi}\Big)^{N / 2} 
e^{-(|\x|^2 + |\y|^2) / 2}
\prod_{j = 1}^{N}(x_jy_j)
\prod_{1 \le j, k \le N}
\Big\{(x_{j}^2 - x_{k}^2)(y_{j}^{2} - y_{k}^{2}) \Big\}
\nonumber\\
& &\times \sum_{\mu:\ell(\mu) \le N}\prod_{j=1}^{N}
\frac{1}{(2\mu_{N-j+1} + 2j -1)!}
s_{\mu}(\{x_{j}^2\})s_{\mu}(\{y_{j}^2\}),
\nonumber
\end{eqnarray}
where 
$\ell(\mu)$ denote the number of parts of the partition $\mu$
(that is, the number of nonzero $\mu_{j}, 1 \le j \le N$). 
Since $s_{\mu}({\bf 0}) = 0$ unless 
$\mu = {\bf 0} \equiv (0, \cdot\cdot\cdot, 0)
\in {\N}_{0}^{N}$, and $s_{{\bf 0}}({\bf 0}) = 1$, 
the asymptotics (\ref{eqn:asymq1}) is obtained.

Next we consider
\begin{eqnarray}
&& q_h^{(N)}(1, \y|\x)
= \left(\frac{2}{h} \right)^N
\sum_{\n \in {\N}^{N}}
\exp \left( - \frac{\pi^2}{2h^2} |\n|^2 \right)
\det_{1 \leq j, k \leq N}
\left[ \sin \left(\frac{\pi}{h} n_j y_j \right)
\sin \left( \frac{\pi}{h} n_j x_k \right) \right]
\nonumber\\
&& = \left(\frac{2}{h} \right)^N
\sum_{\n \in {\N}^{N}}
\exp \left( - \frac{\pi^2}{2h^2} |\n|^2 \right)
\frac{1}{N!} \sum_{\sigma \in S_{N}}
\det_{1 \leq j, k \leq N}
\left[ \sin \left(\frac{\pi}{h} n_{\sigma(j)} y_j \right)
\sin \left( \frac{\pi}{h} n_{\sigma(j)} x_k \right) \right]
\nonumber\\
&& = \frac{1}{N!}
\left(\frac{2}{h} \right)^{N}
\sum_{\n \in {\N}^N}
\exp \left(-\frac{\pi^2}{2h^2}
|\n|^2 \right)
\det_{1 \leq j, k \leq N}
\left[ \sin \left( \frac{\pi}{h} y_j n_k \right) \right]
\det_{1 \leq j, k \leq N}
\left[ \sin \left( \frac{\pi}{h} x_j n_k \right) \right],
\nonumber
\end{eqnarray}
where $\N=\{1,2, \cdots\}$
and $S_N$ is the set of all permutations of 
$N$ items $\{1,2, \dots, N\}$.
Here
\begin{eqnarray}
&& \det_{1 \leq j, k \leq N}
\left[ \sin \left( \frac{\pi}{h} x_j n_k \right) \right]
\nonumber\\
&=& \det_{1 \leq j, k \leq N} \left[
\sum_{m=0}^{\infty} \frac{(-1)^{m}}{(2m+1)!}
\left(\frac{\pi}{h} x_j n_k \right)^{2m+1} \right]
\nonumber\\
&=& \sum_{\m \in {\N}_0^{N}}
\prod_{j=1}^{N} \left\{
\frac{(-1)^{m_j}}{(2m_j+1)!} x_j n_j \right\}
\left( \frac{\pi}{h} \right)^{2\sum_{j=1}^{N}m_j+N}
\det_{1 \leq j, k \leq N}
\Big[ (x_j n_k)^{2m_j} \Big]
\nonumber\\
&=& \prod_{j=1}^{N}(x_j n_j) 
\sum_{\m \in {\N}_0^{N}}
\left( \frac{\pi}{h} \right)^{2\sum_{j=1}^{N}m_j+N}
\prod_{j=1}^{N}
\frac{(-1)^{m_j}}{(2m_j+1)!} 
\frac{1}{N!} \det_{1 \leq j, k \leq N} [x_j^{2m_k}]
\det_{1 \leq j, k \leq N} [n_j^{2m_k}].
\nonumber
\end{eqnarray}
Therefore
\begin{eqnarray}
&& q_h^{(N)}(1, \y|\x) \nonumber\\
&=& \frac{1}{N!} \left(\frac{2}{h}\right)^{N}
\prod_{j=1}^{N}(x_j y_j)
\sum_{\n \in {\N}^{N}} \exp \left(
-\frac{\pi^2}{2 h^2} |\n|^2 \right)
\prod_{j=1}^{N} n_j^2
\nonumber\\
&& \times \sum_{0 \leq m_1 < m_2 < \cdots < m_N}
\left( \frac{\pi}{h} \right)
^{2 \sum_{j=1}^{N} m_j+N}
\prod_{j=1}^{N} \frac{(-1)^{m_j}}{(2m_j+1)!}
\det_{1 \leq j, k \leq N}[x_j^{2m_k}]
\det_{1 \leq j, k \leq N}[n_j^{2m_k}]
\nonumber\\
&& \times \sum_{0 \leq \ell_1 < \ell_2 
< \cdots < \ell_N}
\left( \frac{\pi}{h} \right)
^{2 \sum_{j=1}^{N} \ell_j+N}
\prod_{j=1}^{N} \frac{(-1)^{\ell_j}}{(2\ell_j+1)!}
\det_{1 \leq j, k \leq N}[y_j^{2\ell_k}]
\det_{1 \leq j, k \leq N}[n_j^{2\ell_k}].
\nonumber
\end{eqnarray}
Note that
$\det_{1 \leq j, k \leq N}
[y_j^{2 \ell_{k}} ]
=(-1)^{N(N-1)/2}
\det_{1 \leq j, k \leq N}
[y_j^{ 2 \ell_{N-k+1}} ]$ 
and set
$\mu_{k}=m_{N-k+1}-N+k,
\nu_{k}=\ell_{N-k+1}-N+k$.
Then
$2 \sum_{j=1}^{N} m_j + N = 2|\mu|+N^2$
with $|\mu| \equiv \sum_{j=1}^{N}\mu_{j}$,
and we have
\begin{eqnarray}
&& q_{h}^{(N)}(1, \y|\x) \nonumber\\
&=& \frac{1}{N!} \left( \frac{2}{h} \right)^N
\left( \frac{\pi}{h} \right)^{2N^2}
\prod_{j=1}^{N}(x_j y_j)
\prod_{1 \leq j<k \leq N}
\{(x_j^2-x_k^2)(y_j^2-y_k^2)\}
\nonumber\\
&& \quad \times \sum_{n \in {\N}^{N}}
\exp \left(-\frac{\pi^2}{2 h^2} |\n|^2 \right)
\prod_{j=1}^{N} n_j^2
\nonumber\\
&& \quad \times 
\sum_{\mu: \ell(\mu) \leq N}
\sum_{\nu: \ell(\nu) \leq N}
\left(\frac{\pi}{h} \right)^{2(|\mu|+|\nu|)}
\prod_{j=1}^{N} \Big\{
\frac{(-1)^{\mu_j+\nu_j}}
{(2\mu_{N-j+1}+2j-1)! (2\nu_{N-j+1}+2j-1)!} \Big\}
\nonumber\\
&& \qquad \times
\det_{1 \leq j, k \leq N}
[n_j^{2(\mu_k+N-k)}]
\det_{1 \leq j, k \leq N}
[n_j^{2(\nu_k+N-k)}]
s_{\mu}(\{x_j^2\}) s_{\nu}(\{y_j^2\}).
\nonumber
\end{eqnarray}
This gives (\ref{eqn:asymq2}).

\section{Proof of Proposition 2}

From (\ref{eqn:result1}), using (\ref{eqn:Vand})
\begin{eqnarray}
&& \P(H^{(N)}_N \leq h) = c_N h^{-N(2N+1)} 
\frac{1}{N!} \sum_{\n \in {\N}^N}
\exp \left( -\frac{\pi^2}{2 h^2} |\n|^2 \right)
\nonumber\\
&& \qquad \times
\prod_{j=1}^{N} n_j^2
\det_{1 \leq j, k \leq N} \Big[ n_j^{2(k-1)} \Big]
\det_{1 \leq \ell, m \leq N} \Big[
n_{\ell}^{2(m-1)} \Big]
\nonumber\\
&&  \quad = c_N h^{-N(2N+1)} 
\frac{1}{N!} \sum_{\n \in {\N}^N}
\prod_{j=1}^{N} \left\{
n_j^2 \exp \left( -\frac{\pi^2}{2 h^2} n_j^2 \right) \right\}
\nonumber\\
&& \qquad  \times
\sum_{\sigma \in S_N} {\rm sgn}(\sigma)
\sum_{\rho \in S_N} {\rm sgn}(\rho)
\prod_{\ell=1}^{N} n_{\ell}^{2(\sigma(\ell)+\rho(\ell)-2)}
\nonumber\\
&& \quad = c_N h^{-N(2N+1)} 
\frac{1}{N!} 
\sum_{\sigma \in S_N} \sum_{\rho \in S_N} {\rm sgn}(\sigma)
{\rm sgn}(\rho)
\prod_{j=1}^{N} \left\{
\sum_{n_j=1}^{\infty} n_j^{2 \sigma(j)+2\rho(j)-2}
\exp \left( -\frac{\pi^2}{2 h^2} n_j^2 \right) \right\}.
\nonumber
\end{eqnarray}
Then
\begin{eqnarray}
&& \P(H^{(N)}_N \leq h)= c_N h^{-N(2N+1)} 
\frac{1}{N!} 
\sum_{\n \in {\N}^{N}} e^{-\pi^2 |\n|^2/(2 h^2) }
\sum_{\sigma \in S_N} \sum_{\rho \in S_N}
{\rm sgn}(\sigma) {\rm sgn}(\rho)
\prod_{j=1}^{N} n_j^{2\sigma(j)+2\rho(j)-2}
\nonumber\\
&& \qquad = c_N h^{-N(2N+1)} 
\frac{1}{N!} 
\sum_{\n \in {\N}^{N}} e^{-\pi^2 |\n|^2/(2 h^2) }
\sum_{\sigma \in S_N} \sum_{\rho \in S_N}
{\rm sgn}(\sigma) {\rm sgn}(\rho)
\prod_{k=1}^{N} 
n_{\sigma^{-1}(k)}^{2 k+2\rho(\sigma^{-1}(k))-2},
\nonumber
\end{eqnarray}
where we have set $k=\sigma(j)$.
Since the summation
$\sum_{\n \in {\N}^{N}}$ is taken, the above is equal to
\begin{eqnarray}
&& c_N h^{-N(2N+1)} 
\frac{1}{N!} 
\sum_{\n \in {\N}^{N}} e^{-2\pi |\n|^2/(2 h^2) }
\sum_{\sigma \in S_N} \sum_{\rho \in S_N}
{\rm sgn}(\sigma) {\rm sgn}(\rho)
\prod_{k=1}^{N} 
n_{k}^{2 k+2\rho(\sigma^{-1}(k))-2}
\nonumber\\
&=& c_N h^{-N(2N+1)} 
\frac{1}{N!} 
\sum_{\n \in {\N}^{N}} e^{-\pi^2 |\n|^2/(2 h^2) }
\sum_{\sigma \in S_N} \sum_{\tau \in S_N}
{\rm sgn}(\tau)
\prod_{k=1}^{N} 
n_{k}^{2 k+2\tau(k)-2}
\nonumber\\
&=& c_N h^{-N(2N+1)} 
\sum_{\n \in \N^{N}} e^{-\pi^2 |\n|^2/(2 h^2) }
\sum_{\tau \in S_N}
{\rm sgn}(\tau)
\prod_{j=1}^{N} 
n_{j}^{2 j+2 \tau(j)-2}
\nonumber\\
&=& c_N h^{-N(2N+1)} 
\sum_{\n \in \N^{N}} e^{-\pi^2 |\n|^2/(2 h^2) }
\det_{1 \leq j, k \leq N}
\Big[ n_j^{2j+2k-2} \Big]
\nonumber\\
&=& c_N h^{-N(2N+1)} 
\det_{1 \leq j, k \leq N}
\left[ \sum_{n=1}^{\infty}
n^{2j+2k-2} e^{-\pi^2 n^2/(2 h^2)} \right],
\nonumber
\end{eqnarray}
where $\tau=\rho \circ \sigma^{-1}$
and the relation
${\rm sgn}(\sigma) {\rm sgn}(\rho)
={\rm sgn}(\tau)$ was used.
Then the first equality of (\ref{eqn:result2})
is proved. It is easy to confirm the second equality.


\newpage 

\end{document}